# Insights on heterogeneity in blinking mechanisms and non-ergodicity using sub-ensemble statistical analysis of single quantum-dots

*Amitrajit Mukherjee[1], Korak Kumar Ray[1], Chinmay Phadnis[1], Arunasish Layek[1], Soumya Bera[2], and Arindam Chowdhury[1]\**

[1]Department of Chemistry, Indian Institute of Technology Bombay, Powai, Mumbai, India.
[2]Department of Physics, Indian Institute of Technology Bombay, Powai, Mumbai, India.

*Email: *arindam@chem.iitb.ac.in*

## Abstract

Photo-luminescence (P-L) intermittency (or blinking) in semiconductor nanocrystals (NCs), a phenomenon ubiquitous to single-emitters, is generally considered to be temporally random intensity fluctuations between '*bright*' ('*On*') and '*dark*' ('*Off*') states. However, individual quantum-dots (QDs) rarely exhibit such telegraphic signal, and yet, the vast majority of single-NC blinking data are analyzed using a single fixed threshold (*FT*) which generates binary trajectories. Further, blinking dynamics can vary dramatically over NCs in the ensemble, and it is unclear whether the exponents ($m_{On/Off}$) of single-particle *On-/Off*-time distributions ($P(t_{On/Off})$), which are used to validate mechanistic models of blinking, are narrowly distributed. Here, we sub-classify an ensemble based on the emissivity of QDs, and subsequently compare the (sub)ensembles' behaviors. To achieve this, we analyzed a large number (>1000) of blinking trajectories for a model system, $Mn^{+2}$ doped ZnCdS QDs, which exhibits diverse blinking dynamics. An intensity histogram dependent thresholding method allowed us to construct distributions of relevant blinking parameters (such as, $m_{On/Off}$). Interestingly, we find that single QD $P(t_{On/Off})$s follow either *truncated power law* or *power law*, and their relative proportion vary over sub-populations. Our results reveal a remarkable variation in $m_{On/Off}$ amongst as well as within sub-ensembles, which implies multiple blinking mechanisms being operational amongst various QDs. We further show that the $m_{On/Off}$ obtained via cumulative single-particle $P(t_{On/Off})$ is distinct from the weighted mean value of all single-particle $m_{On/Off}$, an evidence for the lack of ergodicity. Thus, investigation and analyses of a large number of QDs, albeit for a limited time-span of few decades, is crucial to characterize possible blinking mechanisms or heterogeneity therein.

## I. Introduction

Fluorescence intermittency (or blinking) is a photo-induced phenomenon exhibited by single emitters, for example, single molecules[1,2] and semiconductor nanocrystals (NCs) such as quantum dots (QDs).[3-9] Photo-luminescence (P-L) blinking involves discrete switches in emission intensity between a bright ('*On*') and a dark ('*Off*') intensity state at a seemingly random time scale. There has been considerable effort to understand blinking as it restricts the usage of NCs as single photon sources[10] or in optoelectronic devices such as lasers,[11] and complicates interpretation of single particle tracking measurements to probe bio-molecular dynamics.[12-13] To understand the origins, several models have been proposed; NC blinking has been attributed to charging-discharging (Auger ionization-recombination),[8,9,14,15] influence of long lived multiple traps (defect states) with fluctuating energy barriers,[9,16] amongst several others propositions. However, mechanism(s) of NC blinking is still debatable and question remains whether any particular blinking mechanism is exclusively operational for various NCs (of same material) in the ensemble.[17-25]

In general, to analyze QD blinking, each intensity-time trajectory is considered as a quasi-telegraphic signal between two states ('*On*' or '*Off*') with respect to a single fixed threshold (*FT*) value. Simulated binary blinking trajectories, constructed via assigning intensities above (or below) the threshold as '*On*' (or '*Off*') and the rest as '*Off*' (or '*On*') levels, are subsequently used to extract relevant blinking parameters, such as, *On-/Off*-time duration distributions ($P(t_{On/Off})$). However, QDs frequently exhibit intermediate '*grey'* states with intensities between '*On*' and '*Off*' levels.[26,27] Consequently, there is a continuous distribution of intensities for a single QD blinking trajectory. Under these situations, two-state analysis of blinking traces using *FT* eludes the '*grey*' states and may not necessarily provide reliable blinking parameters. This is particularly relevant in context of various complementary theoretical models for origin(s) of NC blinking, where experimentally extracted properties, such as $P(t_{On/Off})$, are compared to validate proposed mechanisms.

It is well known that single NC $P(t_{On/Off})$ follows either power law (*PL*) or truncated power law (*TPL*) depending on the nature of the system,[6,7,9] with different exponent ($m_{On/Off}$) values and truncation times ($\tau_c$). In several reports,[6,17,28] an average $m_{On/Off}$ value ($\langle m_{On/Off} \rangle$) considering several tens of QDs, has been used to validate certain blinking models. On the other hand, assuming ergodic behavior, the exponent from a single NC $P(t_{On/Off})$ over very long time periods (5-6 decades) has been related with different model predictions.[29] It is believed that '$m_{Off}$' values around 1.5 (or slightly higher) can be expected for a '*Charging-discharging*' model,[8,17,19,30] while '*trapping de-trapping*' processes of charges in surface defects should yield $m_{On/Off}$ in a range from 1 to 2.[6,9,15,31] Alternatively, '*diffusion-controlled electron transfer*'[19,20] predicts that initial exponent values range from 0.5 to 1.5. In contrast however, a wide range of experimental $m_{On/Off}$ (typically between 0.8 and 2.2) have been reported for various NC systems, [6,9,15-17,31-37] which makes it challenging to establish a unique mechanism for intermittency.

In general, one of the above blinking mechanisms is considered to be operational for QDs of the same material composition and similar morphology. However, unlike molecules, every NC is inherently distinct with different number of atoms, surface ligand density, as well as population, nature and spatial distribution of defects.[15] Therefore, depending on the particular QD being probed, the origin of intermittency may vary which should be reflected in single-particle $m_{On/Off}$ values. Interestingly, few recent reports have proposed a possibility of two competing blinking mechanisms being operational simultaneously even for individual QDs.[38-41] This leads to the question of whether the ensemble average of the exponents and the exponent obtained from the time-ensemble averaged $P(t_{On/Off})$ are the same. If they were not, it might imply the loss of ergodicity in blinking processes. This can only be deciphered if blinking parameters are investigated for a large number of QDs, and especially relevant when the intermittency characteristics of various NCs are visibly diverse.

Although intermittency of any two QDs in an ensemble is never identical, even when data is acquired under identical conditions, there are often prominent and striking differences in blinking characteristics of individual emitters. For instance, the blinking dynamics of $Mn^{+2}$ doped ZnCdS QDs[42] with very similar morphology, immobilized in a polymer film, exhibit diverse intermittency patterns in terms of their emissive propensity. Fig. 1(a) shows representative traces from three such individual QDs, being either mostly, moderately, or rarely emissive over the duration of data acquisition. Several other NC systems,[29,43-48] including CdSe-ZnSe alloy core-shell QDs[49] of similar size (Fig. 1(b)) also exhibit such contrasting blinking characteristics in terms of time-averaged intensity distributions, albeit the fraction of different sub-populations may vary. Interestingly however, whether the nature of intermittency have any correlation with blinking parameters (such as $m_{On/Off}$) still remains obscure due to lack of statistically relevant distributions.

Here, we have sub-classified a large number of QDs in the ensemble based on their individual intensity distributions (emissivity) and subsequently evaluated the distributions of blinking parameters. We have chosen $Mn^{+2}$ doped ZnCdS and CdSe-ZnSe alloy core-shell QDs as model systems as both these NCs exhibit diverse blinking propensity in terms of their emissivity, however, we report the results of one of these systems, namely the doped ZnCdS QDs. To address blinking dynamics with '*On*', '*Off*' and '*grey*' intensity levels, here, we have developed an intensity histogram dependent thresholding (*IHDT*) method, where multiple thresholding explicitly considers the presence of '*grey*' states, and the threshold values are based on the intensity distribution of each emitter. For sub-ensemble analysis, we classified the QDs in three broad categories based on the emissivity (% *On*-time, $\tau_{ON}$ (%)) for each QD, namely, "*Mostly Off*", "*Intermediate*" and "*Mostly On*". To understand how blinking parameters vary over each sub-ensemble, we have performed statistical analysis considering more than 1000 single emitters, which is an order of magnitude high than that done in any prior report. This allowed us to compare the mean behaviors as well as the variability of blinking parameters of various sub-ensembles with that of the entire population. Our results indicate a substantial variation in nature of the distributions and the average values of several pertinent parameters, which suggest multiple blinking mechanisms being operational for different emitters in the ensemble leading to the breaking down of ergodicity.

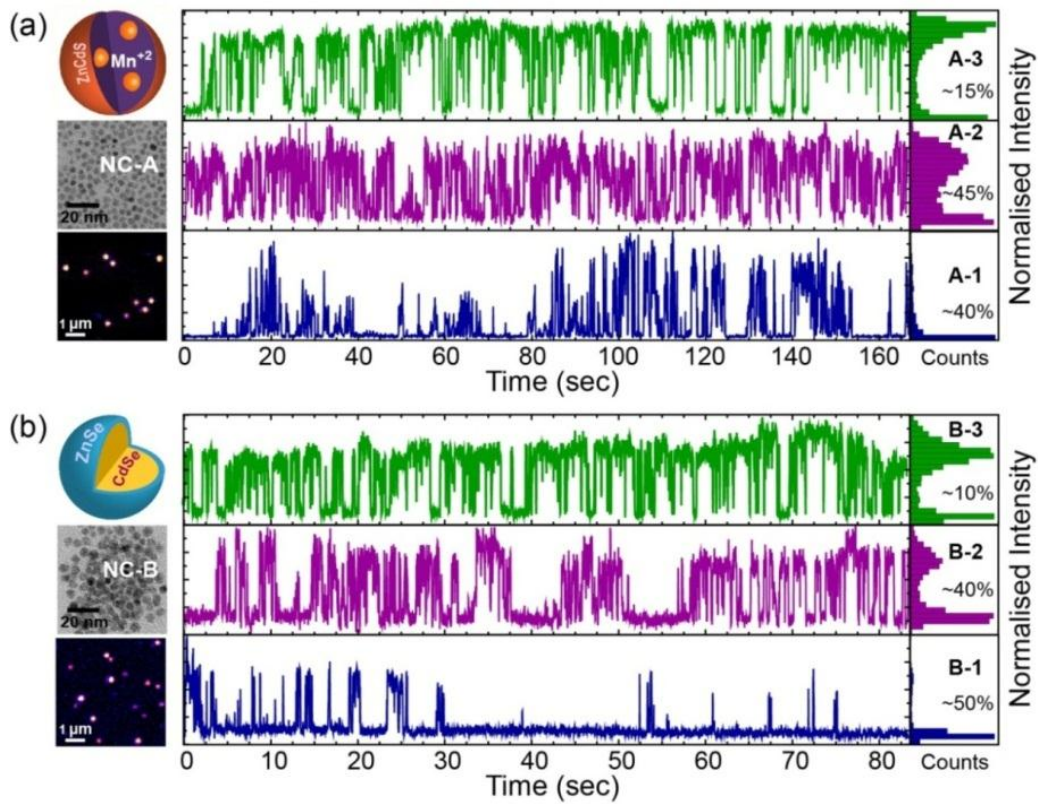

**Figure 1**. Diverse blinking characteristics of two NC systems: (a) $Mn^{+2}$ doped ZnCdS QDs (NC-A) and (b) CdSe-ZnSe alloy core-shell QDs (NC-B), measured under identical conditions (*see* Methods). The middle-left panels are TEM images of NCs depicting uniform size distributions, and lower-left panels are fluorescence images of spatially-segregated NCs. The intermittency behaviors and corresponding time-averaged intensity distributions of the three representative NCs for each system (A1-A3; B1-B3) depict different sub-populations having distinct blinking patterns (color coded), such as being *"Dominantly emissive" (green)*, *"Moderately emissive"(purple),* and *"Mostly non-emissive" (blue)*, as well as frequent occurrence of intermediate '*grey*' states. Such diverse intermittency patterns of a few QDs in the ensemble is shown in Movie M1 (supplementary material).

## II. Experimental Methods

*A. Sample preparation and data acquisition*

$Mn^{+2}$ doped ZnCdS QDs (NC-A) were synthesized and characterized by Abhijit Hazarika as done in Ref. 42, and these samples were provided by Prof. D.D. Sarma. CdSe-ZnSe alloy core-shell NCs (NC-B) were prepared following previously reported literature. Dilute (~nM) solution of both QDs were immobilized in a 300 nm thick polymer film (polymethyl methacrylate, PMMA) via spin coating (2000 rpm) out of chloroform on a freshly cleaned glass coverslip. For details, see supplementary material, and Refs. 42 and 49. The NC-A and NC-B samples were excited by a 457 nm (Argon Ion) cw laser excitation (at 0.5 kW cm$^{-2}$) using a home-built epifluorescence microscopy set up capable of single-molecule detection. The P-L emission from single QDs were passed through appropriate dichroic and emission filters (545-635 nm band-pass for NC-A and 500 nm long pass for NC-B) and detected using an interline CCD camera (DVC 1412AM) where movies (16 bit images) were collected using at 20 Hz. Due to limitation of data collection computer memory, wide field (diameter ~ 30 μm) movies were recorded for 166.5 s which contained 3333 frames (for the NC-A). Each movie containing several tens of single QDs was analyzed after background flattening using ImageJ (*see* supplementary material). The details of the experimental setup can be found elsewhere[42] and specifics of image analyses are provided in supplementary material. All data were collected at 295K.

*B. Data analysis procedures*

To obtain a large number of blinking trajectories in an efficient manner, we have developed a simple MATLAB code, which extracts blinking traces from different QDs within a specified region of interest. Initially, the (*x,y*) coordinates of the maximum intensity positions in the time-averaged image (*i.e.*, from each QD) are extracted using *ImageJ*. Thereafter, the code extracts P-L traces from those (*x,y*) coordinates of the *maximum intensities* for each QDs in a movie, without any manual intervention. Although we have analyzed the blinking from both the samples (NC-A and NC-B), we discuss the results on the $Mn^{+2}$ doped ZnCdS NC system (NC-A) for which the blinking propensity was observed to be relatively more diverse. For the construction of *On-/Off*-time distributions ($P(t_{On/Off})$) we chose only those blinking traces which possess at least five distinct '*On*' or '*Off*' time durations ($t_{On/Off}$), following a least square method with a statistical weightage scheme as done by Kuno.*et al.*[6] Power law (*PL*: $P(t) = A.t^{-m}$) and truncated power law (*TPL*: $P(t) = A.t^{-m}.\exp(-t/\tau_c)$) type $P(t_{On/Off})$ are then segregated according to the magnitude of truncation time ($\tau_c$): an emitter with higher $\tau_c$ value than 10 sec has been observed to follow *PL* nature as a best fit in least square method. Exponent values ($m_{On/Off}$) of $P(t_{On/Off})$ from the single QDs which follow either *PL* or *TPL* have been extracted for further analysis. More details of data analysis procedures are provided in supplementary material.

## III. Intensity Histogram Dependent Thresholding (*IHDT*)

It is known that the choice of the single threshold can lead to artifacts in simulated binary trajectories.[4,24] In some reports, the average noise has been used to set a single *FT*, for instance, the mean of 2-3 times the standard deviation of background counts[6,9,50] or the threshold is chosen as the highest background count over the entire duration of the experiment.[51] An alternate approach has been to choose the FT value from the minimum of the time-averaged (bimodal) intensity distributions via fitting two Gaussians for the '*On*' and '*Off*' states.[52] Regardless of the particular method used to generate the threshold, the majority of reported data use a *single FT* for *all* the QDs to generate binary trajectories for further analyses. However, such a procedure may not be ideally suited for every trajectory in the ensemble especially when the blinking dynamics are heterogeneous and occurrence of "*grey*" levels is common (*see* Fig. 1). For example, we find that the blinking or switching frequency (*SF*) as well as the $P(t_{On/Off})$ change systematically with the choice of *FT* (*vide infra*), owing primarily to miscounting of blinking events (Figs. S1 & S2 in the supplementary material). Therefore, we have developed a method that yields flexible (single or multiple) threshold(s) based on the time-averaged intensity distributions for each QD blinking trajectory, the salient features of which are as described below.

In the multiple thresholding (*IHDT*) approach, we first fit the time-averaged intensity distribution of the entire blinking trajectory using two Gaussian functions. Based on the mean positions of the peaks and the standard deviation of the Gaussians, two values of threshold are chosen (*see* Fig. S3, and methods, supplementary material). The threshold at lower intensity ($I_{th}(On)$) value has been considered as the '*On*' threshold when the change in *PL* intensity between successive frames ($I_t - I_{t-1}$) is positive, while the '*Off*' threshold is set at a higher intensity ($I_{th}(Off)$) when ($I_t - I_{t-1}$) is negative (Fig. 2, and Fig. S3, supplementary material). However, under certain circumstances, such as very dominant "*Mostly On*" or "*Mostly Off*" type blinking trajectories, the algorithm (Figs. S4, supplementary material) allows two thresholds to merge and converge into one single threshold. In case of multiple (two) thresholds, any intensity fluctuations in the region between $I_{th}(On)$ and $I_{th}(Off)$ with duration of more than two consecutive frames (>100 ms), is considered as the '*grey*' or third intensity state (with a value set at 0.5) on top of the '*On*' and '*Off*' states (with values of 1 and 0, respectively). It is important to note that the program also allows us to exclude these '*grey*' (third) intensity states to perform *two*-state blinking analysis, rather than considering an '*On*'-'*grey*'-'*Off*' (three state) scenario. Such detection followed by removal of '*grey*' state without compromising on the blinking events, is exemplified in Fig. 2(b) for one NC. Here, the difference between *FT* (black dotted line) and *IHDT* (red solid line) simulated trajectories can be readily identified in the temporally blown up section of duration ~16 sec. In all subsequent analysis for traces using multiple thresholds generated by *IHDT*, $\tau_{ON}$ (%)s for the individual emitters were calculated after the elimination of the '*grey*' intensity level. More details on the algorithm of the *IHDT* are provided in supplementary material (*see* Figs. S3-S5). The comparison of several blinking parameters obtained using *IHDT* and *FT* methods are discussed in the following section.

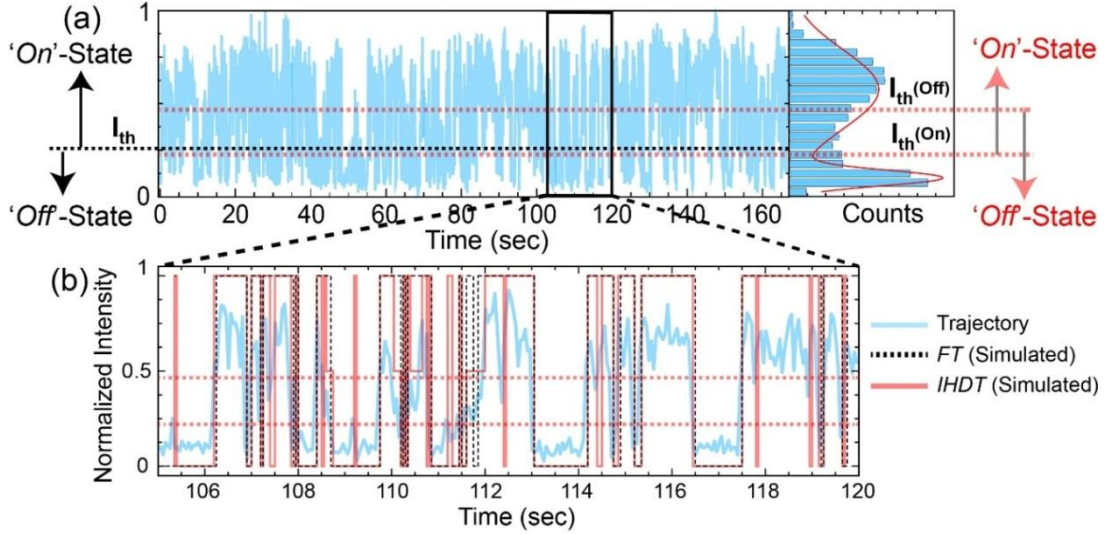

**Figure 2**. The concept of multiple thresholdoing as used by the *IHDT* method and comparison with *FT* (@ 0.25). (a) Normalized intensity trajectory for a single blinking NC (NC-A-2), the corresponding time-averaged intensity distribution and fit to a double gaussian function. The dashed horizontal lines represent single or double thresholds for *FT* and *IHDT* (b) Temporal blowup of a section of the actual trajectory along with the respective simulated trajectories obtained from *IHDT* (red solid line) and *FT* method (black dotted line). Here, in contrast to *FT*, the simulated trajectory from *IHDT* detects the presence of '*grey*' states.

## IV. Results and Discussion

We first compare the results from a *FT* analysis with that for the *IHDT*, and subsequently discuss the results from sub-ensemble statistical analysis. We find that for $Mn^{+2}$ doped ZnCdS QDs (Fig. 1(a), NC-A), the emissivity, $\tau_{ON}$ (%) over 166.5 sec vary between 2.58% and 79%. Therefore, we divided 1040 QDs into three rudimentary sub-ensembles based on equal range of $\tau_{ON}$ (%) (*see* Fig. S6, supplementary material). These three broad sub-ensembles are "*Rarely emissive*" or "*Mostly Off*" ($\tau_{ON}$ (%) < 28%), "*Moderately emissive*" or "*Intermediate*" (28% < $\tau_{ON}$ (%) < 53.5%) and "*Mostly emissive*" or "*Mostly On*" ($\tau_{ON}$ (%) > 53.5%), which are designated as *Categories I, II* and *III*, respectively.

*A. Comparison of IHDT and FT analysis for (sub)ensemble blinking behaviors*

Fig. 3 shows the nature of the $P(t_{On/Off})$ and corresponding exponent values ($m_{On/Off}$) of three representative single QD traces from the above-mentioned blinking categories, constructed using a *FT* @ 0.25 (a-c) and the *IHDT* (d-f) method. First, our results from both the *FT* and *IHDT* model reveal that, $P(t_{On/Off})$ of individual QDs can exhibit either *PL* or *TPL* nature (over three decades), which contradicts the notion that QDs of same material composition and size follow very similar blinking statistics. Further, we find that depending on the particular QD being investigated, often there are considerable deviations in exponent values ($m_{On/Off}$) and truncation times ($\tau_c$), and these differences are pronounced for certain categories (sub-ensembles) (*vide-infra*). Our observations indicate that analysis of blinking statistics of individual QDs in the ensemble is likely to produce ambiguous values of blinking parameters, which in turn can lead to difficulty in their interpretation.

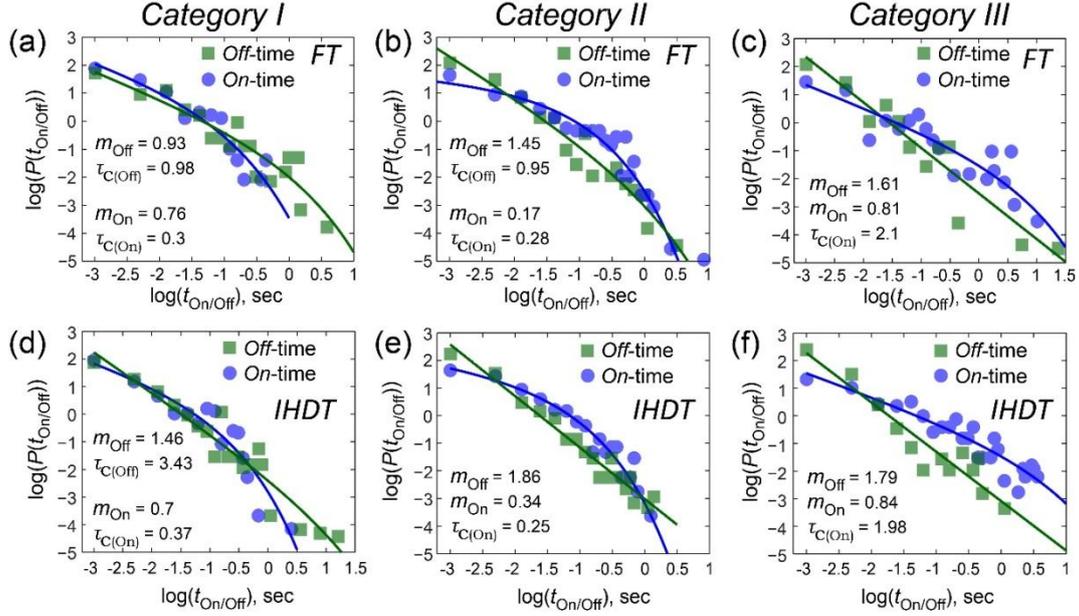

**Figure 3**. Comparison of the $P(t_{On})$ (circles) and $P(t_{Off})$ (squares) obtained using conventional *FT* (@ 0.25) (a-c) and the *IHDT* (d-f) methods, for three single QDs belonging to the sub-ensemble categories *"Mostly Off"* (a,d), *"Intermediate"* (b,e) and *"Mostly On"* (c,f). The experimental data (symbols) extracted using *FT* and *IHDT* follows power law (*PL*) or truncated power law (*TPL*) as exemplified by the fits (solid lines).

With this in mind, we initially analyzed a statistically relevant number (>1000) of $Mn^{+2}$ doped ZnCdS single QDs' blinking trajectories, and compared their rudimentary blinking behaviors, such as the percent $\tau_{ON}$ (%) and switching ('*On*'→'*Off*' or '*Off*'→'*On*') frequency (*SF*) of the ensemble, using *FT* and *IHDT*. The ensemble distributions obtained for the $\tau_{ON}$ (%) and *SF* evaluated using these two methods are shown in Fig. 4. We find that the distributions of both the parameters vary significantly with the chosen threshold values (0.25, 0.4, and 0.55) for the *FT* analyses, and specifically, the mean as well as modal values progressively increase for *SF* with deceasing *FT* (as shown in Fig. 4). Further, the *IHDT* method yields even higher average values for both the *SF* and $\tau_{ON}$ (%) distributions in comparison with *FT* analyses. Clearly, *IHDT* is able to identify a larger number of switching events for individual QDs; *FT* analyses are unable to identify these events because of exclusion of relatively low amplitude and short-duration events (flickering) at both higher and lower intensities with respect to the chosen single threshold. It should be noted that excursions between *On-/Off-* and *'grey'* states were ignored in *IHDT*, which implies the evaluated number of switching events represents the lower bound for the *SF*. This exemplifies one advantage of using *IHDT* over *FT* analysis to capture the number of blinking events ('*On*' to '*Off*' or *vice versa*) closer to the actual value.

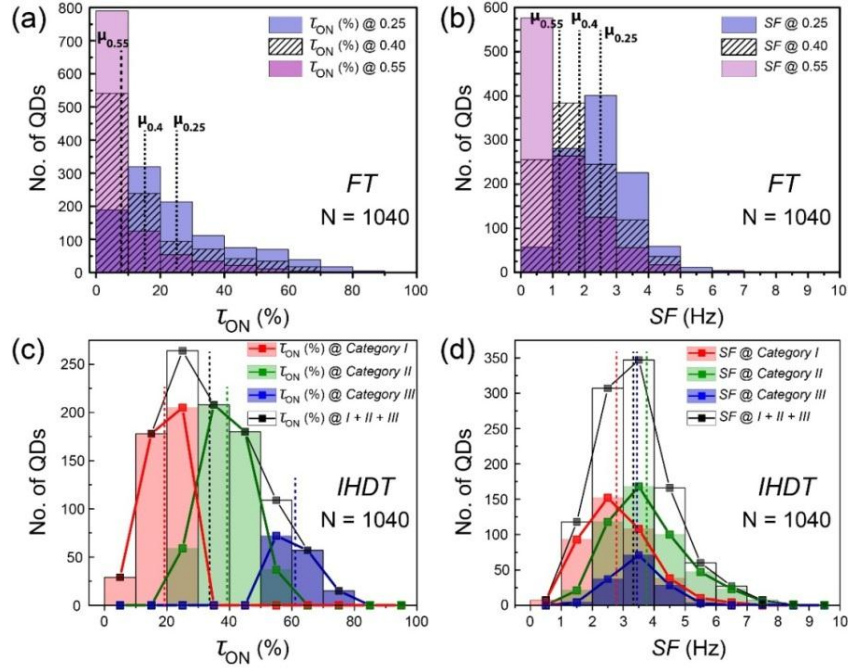

**Figure 4**. Comparison of the distributions of overall percent *On*-time ($\tau_{ON}$ (%)) (a) and switching frequency (*SF*) (b) with the choice of *FT* values at 0.25, 0.4, 0.55. Distributions of $\tau_{ON}$ (%) (c) and *SF* (d) for the entire ensemble (*Categories I + II + III*) (black) as well as that for the three sub-ensembles categories (*I*: red; *II*: green; *III*: purple) obtained using *IHDT*. Ensemble distributions were constructed using 1040 single QDs. Vertical dotted lines represent the mean (μ) for each distribution.

**Table I. (Sub)ensemble $\langle \tau_{ON} (\%) \rangle$ and $\langle SF \rangle$ for *IHDT* and *FT***

| (Sub) Ensemble Category | $\langle \tau_{ON}(\%) \rangle$ IHDT (%) | [a]Δ⟨$\tau_{ON}(\%)$⟩$_{IHDT}$ | | | $\langle SF \rangle$ IHDT (Hz) | [a]Δ⟨$SF$⟩$_{IHDT}$ | | |
|---|---|---|---|---|---|---|---|---|
| | | *FT @ 0.25* | *FT @ 0.40* | *FT @ 0.55* | | *FT @ 0.25* | *FT @ 0.40* | *FT @ 0.55* |
| *I* | 19.41 | +25.3 | +113 | +277 | 2.79 | +24.7 | +80.3 | +182 |
| *II* | 39.15 | −6.75 | +3 | +18.8 | 3.76 | +14.8 | +16.6 | +18.7 |
| *III* | 61.34 | −7.87 | +0.75 | +13.5 | 3.43 | +36.1 | +21.7 | +5.9 |
| *I + II + III* | 34.40 | +35.3 | +126 | +313 | 3.33 | +33.2 | +81.3 | +175 |

[a.] *Percent change in IHDT with respect to FT*

To decipher the origin of the observed deviations, we compared these blinking parameters of sub-populations with the ensemble. Out of 1040 QDs investigated, 412 (39.62%) were *"Mostly Off"* (*Category I*), whereas 484 (46.54%) and 144 (13.84%) were *"Intermediate" (Category II)*, and *"Mostly On" (Category III)*, respectively. The results obtained from *IHDT* and *FT* on the *mean* $\tau_{ON}$ (%) ($\langle \tau_{ON} (\%) \rangle$) and *average* switching frequency ($\langle SF \rangle$) for the category-wise sub-ensembles are shown in Table I (also *see* Fig. S6, supplementary material). While $\langle \tau_{ON} (\%) \rangle$ increases for *IHDT* as compared to *FT* analysis for the entire ensemble, this behavior is not necessarily true for all the sub-populations, such as for *Category II and III*. For instance, $\langle \tau_{ON} (\%) \rangle$ evaluated using *IHDT* decreases with respect to that for the *FT* analysis. This is because a lower threshold (at 0.25) in *FT* analysis often considers a '*grey*' state as an '*On*' state, which frequently appears in *Category II and III* type emitters, and thus increases the value of $\langle \tau_{ON} (\%) \rangle$. However, for *IHDT* considers '*grey*' states neither as '*On*' nor '*Off*' states, thereby resulting a lower value of $\langle \tau_{ON} (\%) \rangle$ for these sub-ensembles. In contrast, the choice of moderately high value of *FT*s (0.4) is close to the mean of the normalized intensity distribution and falls within the regime of '*grey*' states. As a consequence, *FT* distributes '*grey*' states as either '*On*' or '*Off*' states, which statistically cancels out in evaluation of $\langle \tau_{ON} (\%) \rangle$. In the same note, since *IHDT* does not assign '*grey*' states as '*Off*' states, a higher value of *FT* (0.55) results in lower $\langle \tau_{ON} (\%) \rangle$ for both *Category II and III* QDs.

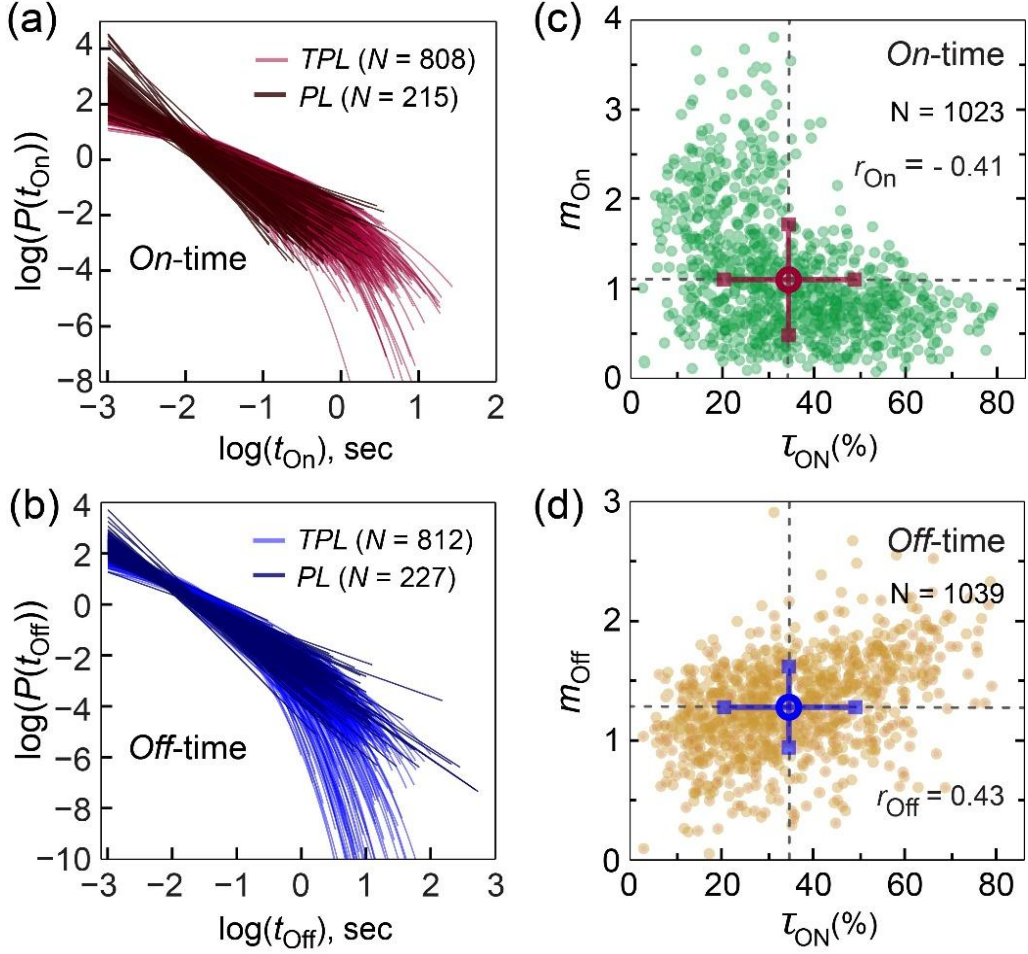

**Figure 5**. The best fits to individual $P(t_{On})$ (a) and $P(t_{Off})$ (b) for more than 1000 single $Mn^{+2}$ doped ZnCdS QDs measured under identical conditions, obtained using *IHDT* method. For both '*On*' and '*Off*' time, darker and lighter hue represent *PL* and *TPL* fits, respectively. The scatter plot of *On*-time (c) and *Off*-time (d) exponents ($m_{On}$ and $m_{Off}$) obtained from individual $P(t_{On})$ and $P(t_{Off})$ as a function of $\tau_{ON}$ (%) for each QD. Circles represent the mean values, while the lengths of horizontal/vertical lines denote twice the value of the standard deviation (σ). Here, $r_{(On/Off)}$ are the Pearson Cross Correlation Coefficients.

*B. Statistical behavior of single-particle $P(t_{On/Off})$ for the (sub)ensemble*

As mentioned earlier in Fig. 3, depending on the particular QD being investigated, the single-particle $P(t_{On/Off})$ either follows *PL* or *TPL*. The fits to experimental $P(t_{On/Off})$ obtained using *IHDT*, for more than 1000 individual $Mn^{+2}$ doped ZnCdS QDs are shown in Figs. 5(a)-5(b). We observe that the majority (~75%) of single NC $P(t_{On/Off})$ has *TPL* nature, and the relative proportion of QDs that exhibit *TPL* (or *PL*) behavior are not severely affected with the mode of analysis (*IHDT or FT @ 0.25*) (Fig. S7 and Table SI, supplementary material). To understand how the $P(t_{On/Off})$ changes with the nature of intermittency, we plotted all extracted $m_{On/Off}$ against the corresponding $\tau_{ON}(\%)$ for each QD (Figs. 5(c)-5(d)). We find that the exponents are very widely distributed ($\sigma_{On} = 0.68$, $\sigma_{Off} = 0.38$) with mean values, $\langle m_{On} \rangle = 1.11$ and $\langle m_{Off} \rangle = 1.28$. It is important to mention that the $m_{On/Off}$ for single NCs can be as low as ~0.1 for quite a few emitters, and as high as ~4 (for *On*-time) and ~3 (for *Off*-time), which is significantly greater than previously reported values (~2). Furthermore, Figs. 5(c)-5(d) depicts either positive or negative correlation ($r_{On} = -0.41$; $r_{Off} =$

0.43) between single NC exponents ($m_{On/Off}$) and the corresponding $\tau_{ON}(\%)$. This implies that, depending on whether they are mostly or rarely emissive, various sub-populations of QDs in the ensemble are likely to have contrasting $m_{On}$ and $m_{Off}$. Effectively, these statistically relevant distributions of $m_{On/Off}$ demonstrates remarkable heterogeneity of blinking dynamics and suggests the possibility of different blinking mechanisms for various QDs (*vide infra*).

To understand the origin(s) of the observed diversity of the extracted exponents, we segregated the QDs which exhibit *PL* and *TPL* nature for $P(t_{On/Off})$. The $m_{On/Off}$ values obtained from *PL* and *TPL* type single-particle $P(t_{On/Off})$ are shown in Figs. 6(a)-6(b), where all the QDs investigated are arranged in the increasing order of $\tau_{ON}(\%)$ (*see* Figs. S8 (a)-S8(b) in the supplementary material for comparison with *IHDT* and *FT*). The frequency histograms of $m_{On}$ and $m_{Off}$, for *PL* and *TPL* nature of $P(t_{On/Off})$, are depicted in Figs. 6(c)-6(d), where the corresponding mean values, $\langle m_{On}\rangle$ and $\langle m_{Off}\rangle$, are represented using vertical lines. We find that for the majority of QDs, $m_{On/Off}$ are close to or below 1.5 for *TPL*, while the corresponding values for *PL* are typically higher, as reflected in their mean values. Moreover, irrespective of *PL* or *TPL* nature of $P(t_{On/Off})$, $P(m_{On})$ is more widely distributed compared to $P(m_{Off})$, more so for turning into a dark state ('*On*'→'*Off*'). It is relevant to mention here that for the QDs which exhibit *TPL* nature of $P(t_{On/Off})$, the (sub)ensemble average truncation times ($\tau_c$) (Table II), which has been related to the probability of charge trapping/recombination processes,[53,54] is nearly half for $P(t_{On})$ than that for $P(t_{Off})$ (~1.3 s). This indicates the occurrence of trapping ('*On*'→'*Off*') for '*On*' events is less frequent than the de-trapping ('*Off*'→'*On*') processes for the '*Off*' events.

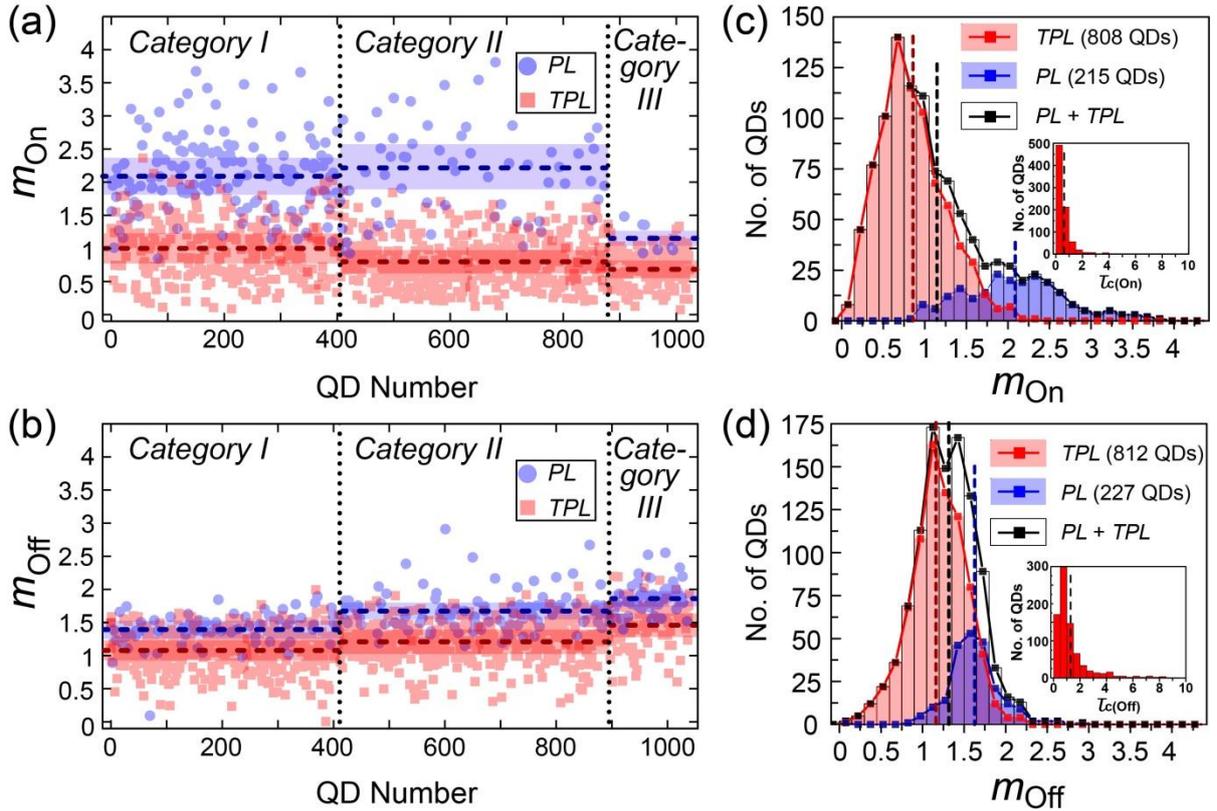

**Figure 6**. $m_{On}$ (a) and $m_{Off}$ (b) obtained from PL (circles) or TPL (squares) nature of $P(t_{On/Off})$, plotted QD number arranged in an increasing order of $\tau_{ON}(\%)$. The horizontal dashed lines represent the $\langle m_{On}\rangle$ and $\langle m_{Off}\rangle$ and shaded region represents the $\sigma(m_{On/Off})$ within each sub-ensemble (for PL and TPL within *I*, *II* and *III*). The histogram of $m_{On}$ (c) and $m_{Off}$(d) for the entire ensemble (black) and that for QDs which exhibit *PL* (purple) and *TPL* (red) nature of $P(t_{On/Off})$. Here, the histograms of truncation times ($\tau_{c\,(On/Off)}$) for *TPL* is shown as insets, and the vertical dashed lines denote the mean values of the distributions.

Sequential arrangement of all the QDs shown in Figs. 6(a)-6(b) further allowed us to compare how *PL/TPL* nature and $m_{On/Off}$ vary amongst various sub-ensembles, (*i.e.*, categories *I*, *II* and *III*). The (sub)ensemble $\langle m_{On} \rangle$ and $\langle m_{Off} \rangle$ (horizontal lines, Figs. 6(a)-6(b)) for each blinking category are shown in Table II. First, our analysis reveals that the proportions of QDs which exhibit *PL/TPL* nature of $P(t_{On/Off})$ vary significantly among the three sub-ensembles (Table II); the ratio of the population of QDs which exhibit *TPL* against *PL* is the highest (lowest) for $P(t_{On})$ for *"Mostly On"* (*"Mostly Off"*) QDs. In contrast however, a reverse trend is observed for $P(t_{Off})$. Further, irrespective of *PL* or *TPL* nature of $P(t_{On/Off})$, the $\langle m_{On} \rangle$ and $\langle m_{Off} \rangle$ (Table II) depends on the (sub)ensemble categories; for instance, $\langle m_{On} \rangle$ decreases along category *I*→*III* sub-populations, while $\langle m_{Off} \rangle$ exhibit an opposing behavior (Figs. 6(a)-6(b), and Table II). Such variation in $\langle m_{On/Off} \rangle$ as well as alteration in the ratio for *TPL* to *PL* characteristic of $P(t_{On/Off})$ among the sub-ensembles reflects diverse relative proportions of "*long*" to "*short*" *On-/Off*-event durations in the blinking trajectories of QDs in the ensemble. Thus, (blinking) category-wise fluctuations further contributes to the overall heterogeneity in $m_{On/Off}$. We emphasize that the deviation of category-wise $\langle m_{On} \rangle$ and $\langle m_{Off} \rangle$ for QDs which exhibit either *PL* or *TPL* $P(t_{On/Off})$ can often be significant, and the $\langle m_{On} \rangle$ and $\langle m_{Off} \rangle$ obtained for the entire ensemble also does not represent the sub-ensemble average behaviors. This provides strong evidence that the various sub-ensemble categories likely originate from different blinking mechanisms.

**Table II. (Sub)ensemble blinking parameters from single-particle $P(t_{On/Off})$**

| (Sub) Ensemble Category | *On*-time[a] | | | | *Off*-time[a] | | | |
|---|---|---|---|---|---|---|---|---|
| | *Power Law Distribution* | *Truncated Power Law distribution* | | | *Power Law Distribution* | *Truncated Power Law distribution* | | |
| | %QDs | $\langle m_{On} \rangle$ | $\langle m_{On} \rangle$ | $\langle \tau_{C(On)} \rangle$[b] | %QDs | $\langle m_{Off} \rangle$ | $\langle m_{Off} \rangle$ | $\langle \tau_{C(Off)} \rangle$[b] |
| *I* | 37 | 2.11 | 0.99 | 0.56 | 17 | 1.40 | 1.08 | 1.39 |
| *II* | 12 | 2.15 | 0.81 | 0.60 | 23 | 1.66 | 1.21 | 1.20 |
| *III* | 5 | 1.15 | 0.70 | 0.75 | 36 | 1.86 | 1.47 | 1.58 |
| *I + II + III* | 21 | 2.09 | 0.85 | 0.61 | 22 | 1.63 | 1.19 | 1.32 |

[a] Weighted (entire ensemble) means of $\langle m_{On} \rangle$ and $\langle m_{Off} \rangle$ combining both *TPL* and *PL* is 1.11 ± 0.68 and 1.28 ± 0.38, respectively. (Sub)-ensemble standard deviations of $m_{On/Off}$ are shown in Table SII.
[b] in seconds

*C. On the ergodicity in blinking process*

In this study, we have considered ergodicity as a uniform blinking mechanism present throughout the ensemble of QDs, which has been predicted in various reports via evaluation of the $m_{On/Off}$ for $P(t_{On/Off})$ from a few tens of QDs.[8,9,14-21] However, there are evidences for nonergodic behaviors in terms of diverse blinking processes, mostly based on the intermittency characteristics of a few NCs[54-56] or a small ensemble of QDs.[34] Our observations on the diversity of single-QD $m_{On/Off}$ for more than 1000 emitters clearly suggest the possibility of various distinct *On-/Off*-mechanisms in the ensemble, in tune with recent reports of simultaneous occurrence of two blinking mechanisms.[39-41] Below, we provide arguments, based on the nature of $P(m_{On/Off})$ and analysis of various sub-ensembles, on the lack of ergodicity for blinking processes.

It is interesting that, apart from being considerably broad, both $P(m_{On})$ and $P(m_{Off})$ over the entire ensemble (Figs. 6(c)-6(d)) are not uniformly distributed (*i.e.*, not symmetric) around their mean values. This skewness arises primarily due to contribution from two distinct nature (*PL* and *TPL*) of $P(t_{On/Off})$. It is important to note that the $P(m_{On/Off})$ independently constructed from *PL* and *TPL* $P(t_{On/Off})$ (Figs. 6(c)-6(d)) are nearly symmetric. Our analysis shows that the skewness in the overall ensemble $P(m_{On/Off})$ originates from the difference in the *PL* and *TPL* distributions' sub-populations. Thus, the larger extent of skewness for $m_{On}$ is a consequence of the significantly higher shift from $\langle m_{On}(TPL)\rangle$ to $\langle m_{On}(PL)\rangle$ values (1.25), compared to that for the $m_{Off}$ (0.44). In addition, we find that the coefficient of variations (COVs) for the sub-ensemble of QDs which exhibit *TPL* nature are quite large, much more so for $m_{On}$ (47%) as compared to $m_{Off}$ (28%). The high COV for $m_{On}$ for *TPL* behavior, which significantly contributes to the skewness of the entire $P(m_{On})$, indicates that the *On*-process ('*Off*'→'*On*') is relatively more uniform than the *Off*-process ('*On*'→'*Off*') among the QDs. This points out that the '*On*' and '*Off*' mechanisms may be different from each other, as suggested in a few earlier reports.[23,57,58] To verify the extent of heterogeneity within the ensemble $m_{On/Off}$ values (*i.e.*, blinking process) we calculated the average of $m_{On}$ and $m_{Off}$ data points (for *PL*, *TPL* and *PL* + *TPL*) *above* ($\langle m_{On/Off}(A)\rangle$) and *below* ($\langle m_{On/Off}(B)\rangle$) the overall mean for each of the distributions. This circumvents binning artifacts of the exponent histogram (Figs. 6(c)-6(d)), and is relevant for relatively small sample sizes. We observe that the deviation of $\langle m_{On/Off}(A)\rangle$ and $\langle m_{On/Off}(B)\rangle$ from the corresponding (sub)ensemble means are significantly different only for $P(m_{On}(TPL))$ and $P(m_{On}(PL + TPL))$ (*see* Fig. S9, supplementary material). This provides evidence on the loss of ergodicity for the overall *Off*-mechanisms ('*On*'→'*Off*').

To substantiate this, it is imperative to compare the ensemble averaged $\langle m_{On/Off}\rangle$ for all the QDs, with the time-averaged exponent ($\overline{m_{On/Off}}$) from individual blinking traces. It is important to note that, extraction of reliable ($\overline{m_{On/Off}}$) requires the analysis of extremely long time (5-6 decades at least) blinking data. However, collection of such long time *PL* data on single QDs is particularly challenging, due to material degradation with prolonged illumination (photo-bleaching) as well as practical limitations such as stage/focus drifts and data acquisition/storage capability. However, it is relatively easy to construct $P(t_{On/Off})$ using hundreds of traces from individual QDs each containing few thousand frames. Therefore, one practical approach has been to combine the $P(t_{On/Off})$ acquired from many blinking trajectories of limited duration (~3 decades), and assume under the ergodicity hypothesis, that the cumulative $P(t_{On/Off})$ would reflect the behavior of one single QD over an extremely long time.[3,17]

Therefore, to test whether such analyses is applicable for QDs, we have constructed time-ensemble averaged (cumulative) *On-/Off*-time distribution ($\langle\overline{P(t_{On/Off})}\rangle$) using *IHDT*. Fig. 7 shows the $\langle\overline{P(t_{On/Off})}\rangle$ and the extracted exponents ($\langle\overline{m_{On/Off}}\rangle$) for the entire ensemble, along with the three sub-ensemble categories. Interestingly, although individual QD may exhibit either *PL* or *TPL* nature for $P(t_{On/Off})$, we find that the $\langle\overline{P(t_{On/Off})}\rangle$ for the (sub)ensemble always follow *TPL*. Such *TPL* nature of $\langle\overline{P(t_{On/Off})}\rangle$ owes to the accumulation of long *On-/Off*-duration events from many emitters and *TPL* nature of $P(t_{On/Off})$ for a dominant fraction (~75%) of QDs. More importantly, the time-ensemble averaged exponents ($\langle\overline{m_{On}}\rangle$ = 1.69, $\langle\overline{m_{Off}}\rangle$ =1.89) of the entire population differs considerably from the respective ensemble (weighted) average (for *PL* + *TPL*) exponent values ($\langle m_{On}\rangle$ = 1.11 and $\langle m_{Off}\rangle$ = 1.28, Figs. 5 and Table II). Further, a significant deviation between the time-ensemble averaged and ensemble averaged values are found for the various sub-ensemble categories (Fig. 7 and Table II); the mismatch between $\langle\overline{m_{On/Off}}\rangle$ and $\langle m_{On/Off}\rangle$ for ensemble and different blinking sub-categories provides additional evidence for a loss of ergodicity in blinking process of QDs.

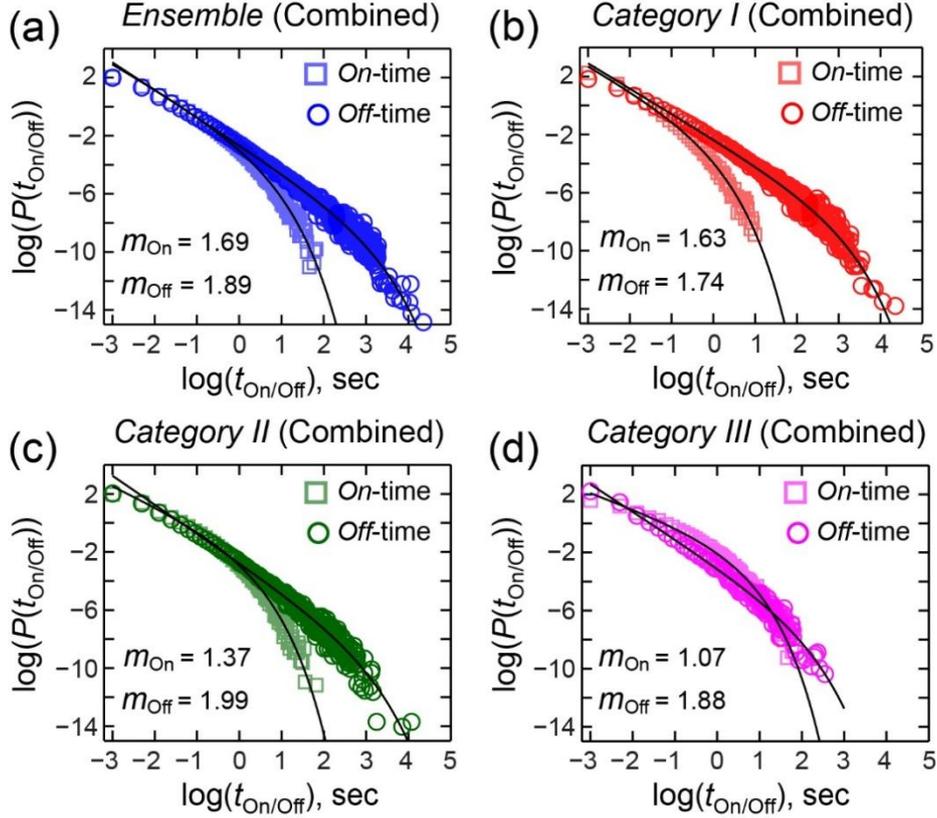

**Figure 7.** Time-ensemble averaged $P(t_{On})$ (squares) and $P(t_{Off})$ (circles) ($\langle\overline{P(t_{On/Off})}\rangle$) for the entire ensemble (a) and that for the sub-ensembles *Category I* (b), *Category II* (c) and *Category III* (d), obtained from combination of single QD $P(t_{On/Off})$. The exponents $\langle\overline{m_{On/Off}}\rangle$ obtained from each of the combined distributions are shown for comparison with the ensemble average values provided in Table II.

To validate the loss of ergodicity due to the existence of different inherent blinking mechanisms and their relation with the blinking propensity, we have performed the analysis of variance (ANOVA) for the category-wise $m_{On/Off}$ values with that for the entire population. We find that the value of F-calculated ($F_{calc}(m_{On}) = 81.54$, $F_{calc}(m_{Off}) = 103.25$) is always greater than F-critical ($F_{cri}(m_{On}) = 3.0044$, $F_{cri}(m_{Off}) = 3.0045$). This corroborates the existence of three distinct sub-ensembles of $m_{On/Off}$ as well, and implies a correlation between different blinking mechanisms and intensity distribution based sub-categorization of QDs. Our inference that multiple blinking mechanisms are operational amongst various QDs is reminiscent of molecular diffusion through heterogeneous media, where, apart from normal Brownian diffusion, certain sub-populations exhibit anomalous diffusion/sub-diffusion or corralled diffusion in passive systems (polymer films and gels), and both super-diffusion and sub-diffusion in active systems (cellular environments). The coexistence of various sub-ensembles can lead to loss of ergodicity,[59] *i.e.*, the time-ensemble average does not strictly correspond with the time-averaged values of certain parameters (within finite timescales of measurement), similar to the situation described here.

## V. Summary and Perspective

Using an intensity histogram based flexible thresholding method and sub-ensemble analysis of a large number of individual $Mn^{+2}$ doped ZnCdS QDs, we show that the blinking dynamics of NCs can often be extremely diverse amongst individual species in the ensemble. Our results demonstrate that due to high NC-dependent variability, it is imperative to perform statistical analysis to estimate blinking parameters (such as $m_{On/Off}$) used to validate or contradict theoretical predictions. We further demonstrate that even a simple classification based on the single-particle intensity distributions (or $\tau_{ON}$ (%)) allow us to distinguish statistical behaviors of sub-populations and the entire ensemble. Our results on $Mn^{+2}$ doped ZnCdS reveal that parameters such as *SF* and $m_{On/Off}$ not only vary over different sub-populations, but even the nature of $P(t_{On/Off})$ in each sub-population is not the same for every QD. Furthermore, $m_{On/Off}$ for individual QDs can often be as low as 0.1 or as high as 3 or 4, considerably more diverse than the prior reported exponent values which lie typically between 1 and 2. This indicates the existence of alternate possible mechanisms for NC blinking than those proposed in the literature. Therefore, generalization of blinking mechanisms based on the characteristics of few tens of QDs is potentially misleading because several underlying processes are likely to be responsible for intermittency of different NCs in the ensemble.

Owing to limited number of blinking events or survival times of QDs, an alternate approach has been to combine the *On-/Off*-time duration distributions from individual intensity trajectories of several tens of QDs, and subsequently extract $m_{On/Off}$ from the cumulative distribution. Here, the underlying assumption is that the ergodic hypothesis holds for blinking in NCs, even though there is evidence to the contrary for single QDs probed over extended time. Our results reveal that there is a significant difference between the exponent ($\langle\overline{m_{On/Off}}\rangle$) of time-ensemble averaged *On-/Off*-time distribution $\langle\overline{P(t_{On/Off})}\rangle$ and the ensemble average of the exponents ($\langle m_{On/Off}\rangle$). This suggests that there is a loss of ergodicity in blinking amongst various QDs in the ensemble, owing primarily to the existence of different sub-populations or processes. Therefore, to infer on mechanism(s), neither is it appropriate to analyze extremely long-time blinking trajectories from a handful of QDs, nor is it reliable to extract $P(t_{On/Off})/m_{On/Off}$ from the cumulative combination of many finite-time blinking traces. Rather, the intermittency of a statistically relevant number of single QDs warrants independent investigation, as semiconductor NCs of very similar composition and morphology are, in reality, distinct entities with their own blinking characteristics.

## Supplementary Material

Supplementary material available with details of experimental and analyses methods, supporting data (Figures S1-S10 and Tables SI-SII), along with a supplementary movie (M1).

## Acknowledgements

AM thanks CSIR (India) for Ph.D. scholarships (09/087(0784)/2013-EMR-I) and CP acknowledges IIT Bombay for post-doctoral fellowship. AC acknowledges financial support (SERB grant no. EMR/2017/004878) from the Department of Science and Technology (Govt. of India) to carry out this work. We thank D. D. Sharma and A. Hazarika for providing QD samples. AC appreciates the support from MNRE (India) aided NCPRE, and CRNTS/SAIF at IIT Bombay for usage of central facility equipment. AM thanks Pronay Biswas for assistance in coding with MATLAB, and Sandip Kar for help with data fitting and analyses.

# Supplementary Material

## Insights on heterogeneity in blinking mechanisms and non-ergodicity using sub-ensemble statistical analysis of single quantum-dots

*Amitrajit Mukherjee[1], Korak Kumar Ray[1], Chinmay Phadnis[1], Arunasish Layek[1], Soumya Bera[2], and Arindam Chowdhury[1]\**

[1]Department of Chemistry, Indian Institute of Technology Bombay, Powai, Mumbai, India.
[2]Department of Physics, Indian Institute of Technology Bombay, Powai, Mumbai, India.

*Email: arindam@chem.iitb.ac.in*

## Supplementary Text

**I. Materials and Methods**

*A. Preparation of $Mn^{+2}$ doped ZnCdS nanocrystals (NC-A):* NC-A has been prepared by *Abhijit Hararika at IISc Bangalore* following a procedure reported by Nag *et al.* [1], which established the utilization of effect of crystal lattice mismatch between the host crystal and dopant materials, on the ease of $Mn^{+2}$ doping. Briefly, for a typical synthesis of $Mn^{+2}$ doped ZnCdS NCs, the reaction mixture consisting of 0.289 mmol of CdO (S.D. Fine-chemicals Limited), 0.097 mmol of ZnO (Leo Chemical), 1 mL of oleic acid (Aldrich), and 10 mL of 1-octadecene (Aldrich) was degassed with nitrogen at 150 $^0$C for 30 min. It was then heated up to 310 $^0$C giving a clear solution. 0.0038 mmol of $Mn(CH_3COO)_2.4H_2O$ in oleyl amine (Aldrich, 1 mL) was injected to the above hot solution. Consequently, a solution of S (1.9 mmol in 1 mL of 1-octadecene) was injected followed by growth at 310 $^0$C for 20 min. All the steps have been carried out in a nitrogen atmosphere, and the product NCs were precipitated and washed repeatedly with 1-butanol. The washed NCs after drying under vacuum, dispersed in toluene for optical measurements.

*B. Preparation of CdSe-ZnSe alloy core-shell nanocrystals (NC-B):* Graded and alloyed $Zn_{1-x}Cd_xSe$ NCs were synthesized by single-pot, high temperature wet chemical route, modifying earlier reported method [2,3,4]. In a typical synthesis, zinc acetate dihydrate ($Zn(Ac)_2$. $2H_2O$, (0.4-x) mM), cadmium acetate dihydrate ($Cd(Ac)_2$. $2H_2O$, x mM), oleic acid (OA, 1.5 ml), and 1-octadecene (ODE, 16 g), were loaded in a three neck flask. The solution was purged with Argon at 120 $^0$C for 1 h, after which it was heated to 300 $^0$C rapidly. Source of selenium (Se, 2.4 mM), mixed with trioctylphosphine (TOP, 2 ml), and ODE (1.5 ml), prepared in a glove box was rapidly injected into reaction mixture at 300 $^0$C. The as-synthesized NCs were annealed for different time duration, in same reaction solution, to produce graded and alloyed NCs. Owing to difference in reactivities of Zn and Cd precursors with TOP-Se complex, graded core/shell structures with Cd-rich core and Zn-rich shells were formed. However, due to same reason homogeneous alloy was formed, when the reaction solution was heated for a longer period (180 mins.).

**II. Details of *IHDT* method and analysis:**

*A. IHDT Model:* Initially, *IHDT* constructs intensity histogram (bin 20) from a blinking time trace and fits it with a double Gaussian function to bear a resemblance with two state intensity distribution. Here, the Gaussian peak with higher intensity count has been considered as the *first peak* and lower intensity count as the *second peak* of the double Gaussian function, which is inbuilt in MATLAB. The peak appearing on the *lower intensity end* corresponds to the intensity fluctuations mainly due to shot noise, and termed as *"noise peak"* whereas the other one at *higher intensity end* is termed as *"higher* *intensity peak"*. We have classified the emitters in terms of positions of these two Gaussian peaks which can be well separated, partially merged or completely overlapped.

In former two situations, multiple (two) thresholds are set. For this purpose, we have optimized the MATLAB program in such a way that the *lower* threshold ($I_{th}(On)$) can be placed just above the *"noise peak"*. Hereafter, the *upper* threshold ($I_{th}(Off)$) has been decided based on the type of intensity histograms discussed above. In case of well separated Gaussian components (Fig. S3 (see *Category II, III*)), $I_{th}(Off)$ has been placed just below the *"higher intensity peak"*. For Gaussians which are partially merged (Fig. S3 (*Category II*)), $I_{th}(Off)$ has been put half sigma lower from the mean of *"higher intensity peak"*.

In case of emitters which are mostly emissive (*"Mostly On"*) or non-emissive (*"Mostly Off"*), the intensities are distributed more towards either *higher intensity end* or *lower intensity end* respectively, where the distribution appears like a narrow peak with a broad tail (Fig. S3 (*Category I*)). For such a distribution, double Gaussian fitting results in a broad *second peak* which penetrates into narrower *first peak*. Under this circumstance, *IHDT* introduces a *single threshold* to the blinking trace either just *below* ('*higher intensity peak*') *or above* ('*noise peak*') the prominent narrower *first peak* of the intensity distribution, depending on whether it is *"Mostly On"* or *"Mostly Off"*. Threshold positions for intensity distribution and blinking trajectories of three categories of emitters are shown in Fig. S3 & Fig. S5. Further, *On-/Off-*time distributions ($P(t_{On/Off})$) for individual single QDs were constructed log-log scale (with base '*e*', ln).

*B. Analysis of Variance (ANOVA):* To check the validity of such intensity distribution based categorization, we have performed Analysis of variance (ANOVA)[5] for the blinking traces in the (sub-)ensemble(s). For this purpose we implemented a statistical F-test on two basic blinking parameters, $\tau_{ON}$ (%) and *SF*, considering individual QDs. In ANOVA, F-critical value extracted from F-table for the three categories within an ensemble of 1040 single emitter QDs has been calculated. Here, the values for degree of freedom (DOF), $df_1$ (DOF between classes), related to the number of sub-categories (conditions) and $df_2$ (DOF within classes), associated with the total population of the ensemble, are found to be 2 and 1037 respectively. The F-critical value (for $df_1$ and $df_2$) for these two DOFs is 3.0044 at confidence level α = 0.05 which describes at least one difference between the sub-ensemble averages ($H_α$). We have considered the null hypothesis ($H_0$) as the equality of the mean values of the three sub-categories ($μ_1 = μ_2 = μ_3$). Further, we have compared the F-calculated and F-critical to ensure the actual existence of three sub-populations if and only if F-calculated becomes greater than F-critical. Results from ANOVA reveals that the calculated F-values, considering two blinking parameters $\tau_{ON}$ (%) and *SF* for each of the QDs, are 2347 and 83.6 respectively. High calculated F-value for $\tau_{ON}$ (%) is expected as the classification of QDs has been performed on the basis of the same blinking property. However, in both of these cases, greater value of F-calculated than F-critical (3.0044) rejects the null hypothesis of equal sub-ensemble averages of blinking properties and therefore provides confidence in sub-categorization of the blinking trajectories for the various QDs in the ensemble.

**Supplementary Data (Figures)**

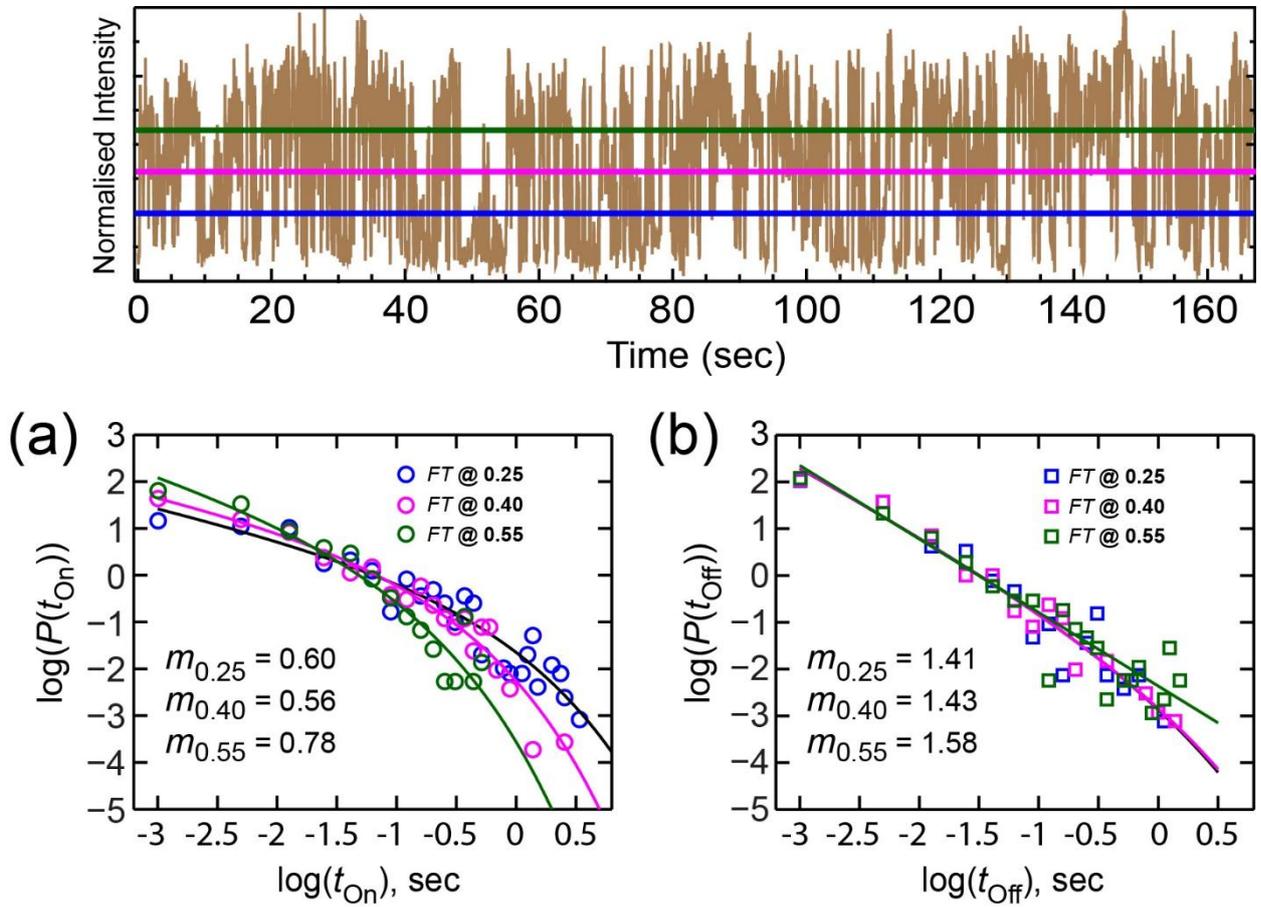

**Figure S1.** The upper panel represents different fixed threshold positions (*FT* @ 0.55 (*Blue*), @ 0.4 (*Magenta*) & @ 0.55 (*Green*)) for NC-A-2 blinking trace. (a) Change in the *On*-time distribution ($P(t_{On})$) with different applied *FT* (@ 0.25, 0.4 & 0.55). (b) Un-identical exponent values as well as the nature (*TPL* to *PL*) of *Off*-time distribution $P(t_{Off})$ with the same thresholds.

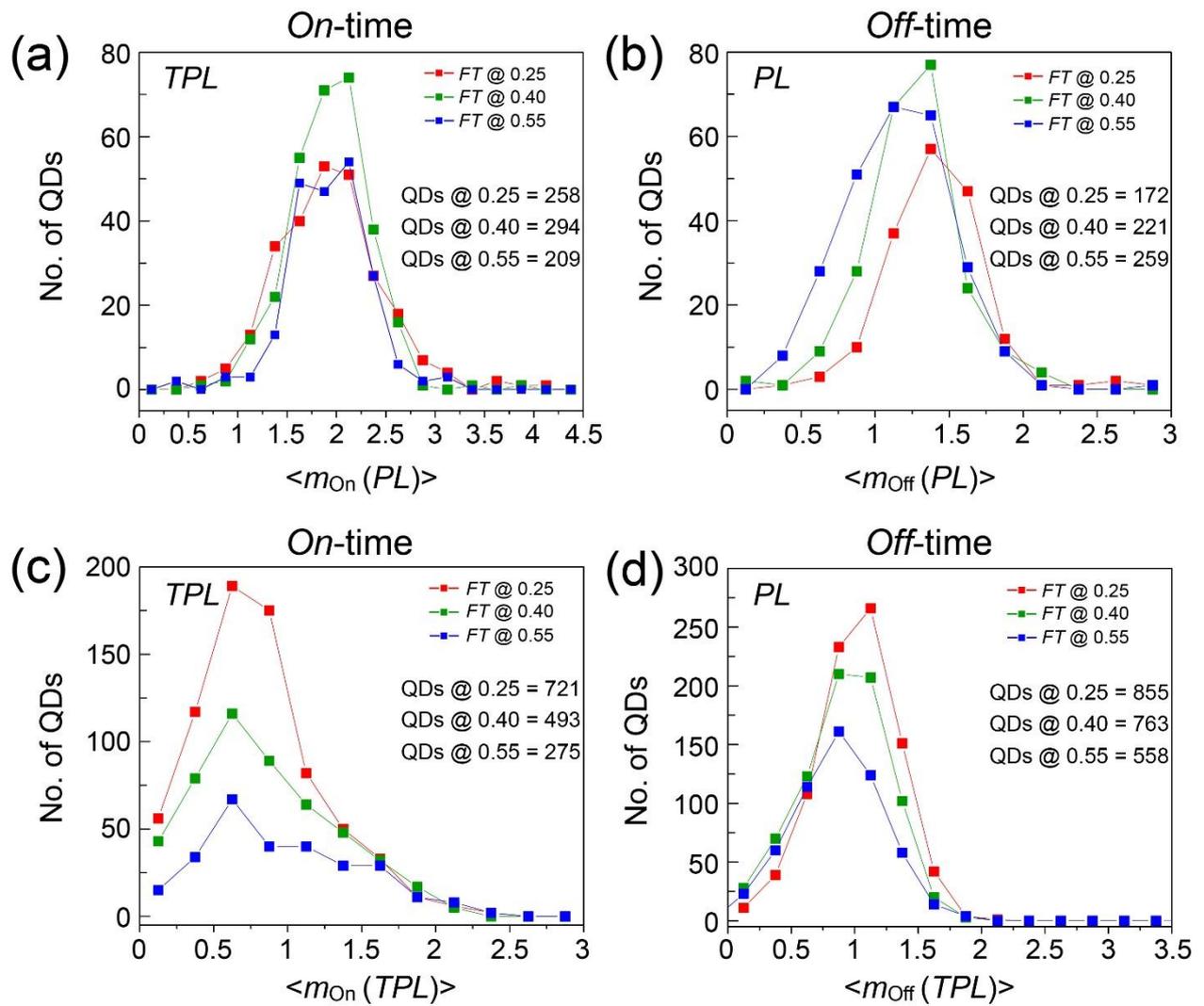

**Figure S2.** Variation in distributions of power law (*PL*) exponents ($m_{On/Off}$ (*PL*)) for *On*- (a) and *Off*-time (b) distribution ($P(t_{On/Off})$) respectively with different *FT* values 0.25, 0.40 and 0.55. Deviation in the distributions of truncated power law (*TPL*) exponents ($m_{On/Off}$ (*TPL*)) for *On*- (c) and *Off*-time (d) distribution, respectively, for varying *FT* threshold values.

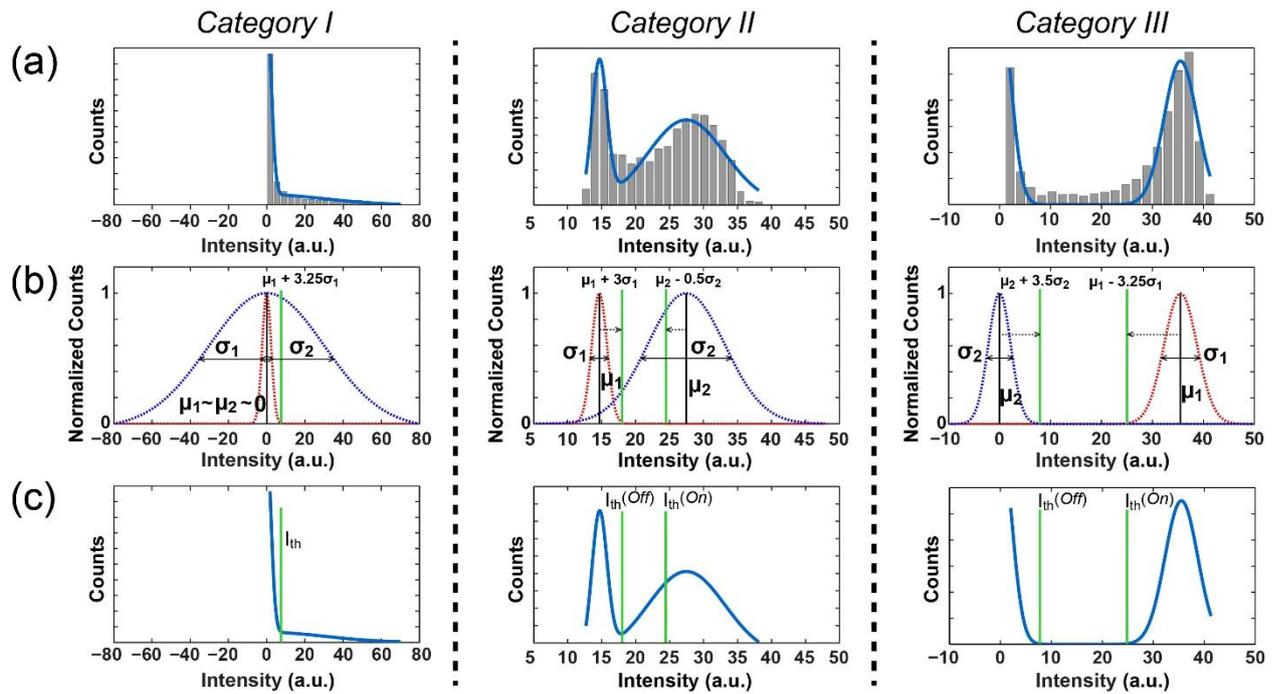

**Figure S3.** Various intensity histograms of single-QD (a), with double gaussian fit function (blue), for three representative *"Mostly Off"*, (*Category I*), *"Intermediate"* (*Category II*) and *"Mostly On"* (*Category III*) emitters. Two normalized gaussian components (blue and red) of the double gaussian functions fitted in the frequency counts of the intensity histograms (b) of emitters. For *Category I* QDs, two gaussians are totally marged one in another while in *Category II* QDs gaussians are partially overlaped and in case of *Category III* QDs these two gaussians are well separated. Gaussians have been normalized to compare the two mean positions easily *T*hreshold ($I_{th}$) position(s) (green vertical line) from *IHDT* for corresponding three representative emitters (c). Intensity axis has been extended toward negative (-ve) axis to show the extrapolated fitting functions for better interpretation.

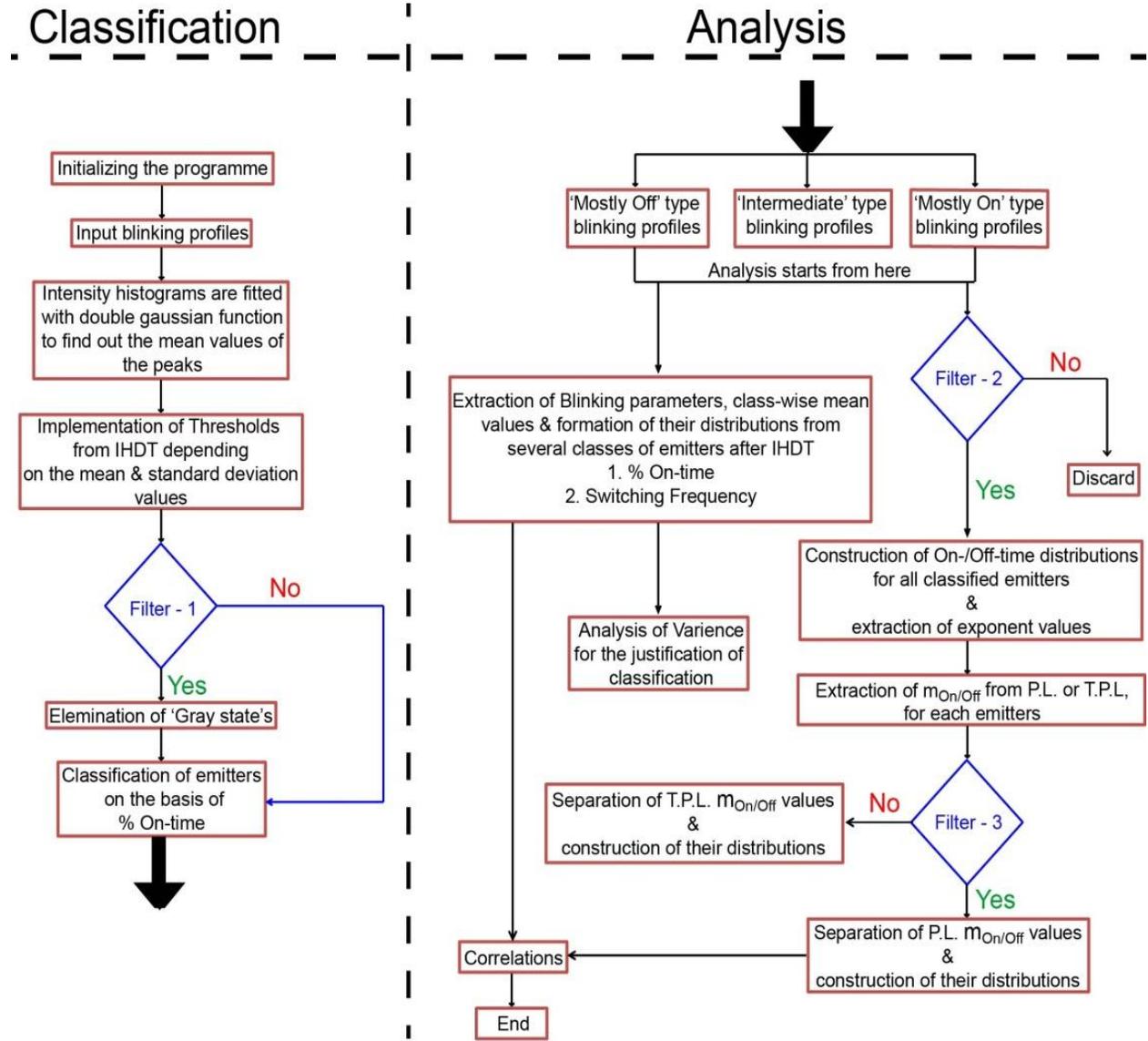

**Figure S4.** This figure represents the *flow chart* of the *IHDT* model, which has mainly two sections, *Classification* and *Analysis* technique. *IHDT* primarily starts segregating the QD blinking trajectories on the basis of corresponding % *On*-time ($\tau_{ON}$ (%)), after successful elimination of '*grey*' states from the trajectory. In the next step these trajectories are analyzed to extract several blinking parameters which will be treated further for (sub)ensemble statistical analysis. Different filters have been used in this algorithm to discard un-efficient blinking traces. **Filter-1** detects the intensities which halt between '*On*' and '*Off*' time thresholds (in case of multiple threshold) for more than 2 consecutive frames (100 ms for NC-A QD) as the '*grey*' state and eliminates them from the two-state analysis. **Filter-2** rejects the blinking traces with less than five *On-/Off*-time duration ($t_{On/Off}$) and discard the trace from construction of $P(t_{On/Off})$. **Filter-3** considers the QD to follow *PL* characteristics of the blinking traces if the truncation (cut-off) time ($\tau_{c\,(On/Off)}$) is greater than 10 seconds, or it is decided to follow *TPL* nature. The *IHDT* analysis technique contains an inbuilt ANOVA check as a part of the MATLAB program to justify the true existence of sub-ensembles within the ensemble.

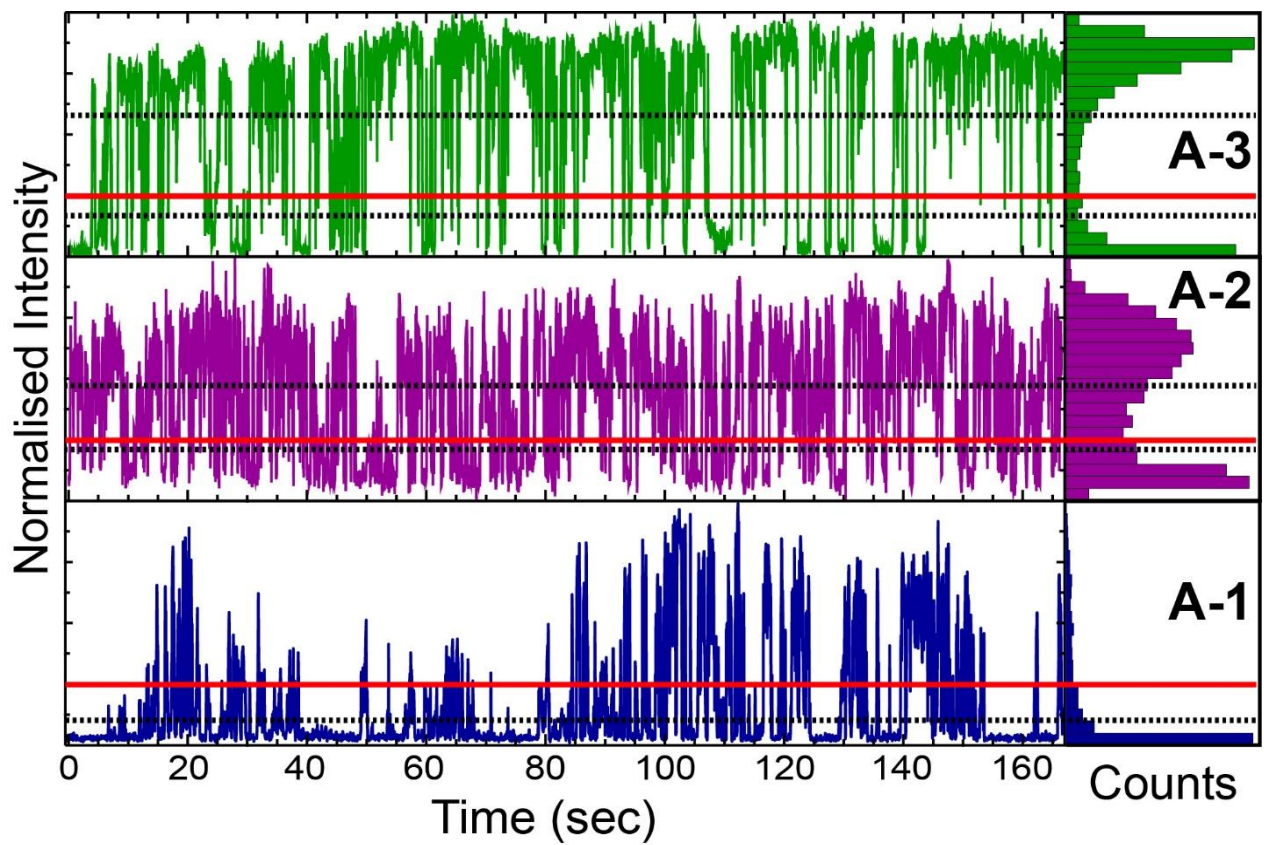

**Figure S5.** Shows threshold positions in fixed threshold (*FT*) (@ 0.25, red lines) and *IHDT* method (black dotted lines) for three normalized representative $Mn^{+2}$ doped ZnCdS (NC-A) single emitter blinking trajectories. Here A-1, A-2 and A-3 represents *rarely emissive* (*Category I*), *moderately emissive* (*Category II*) and *mostly emissive* (*Category III*) QDs. In *IHDT* normalized thresholds for A-1, A-2 and A-3 are 0.11, (0.2245, 0.47) and (0.17, 0.58) respectively.

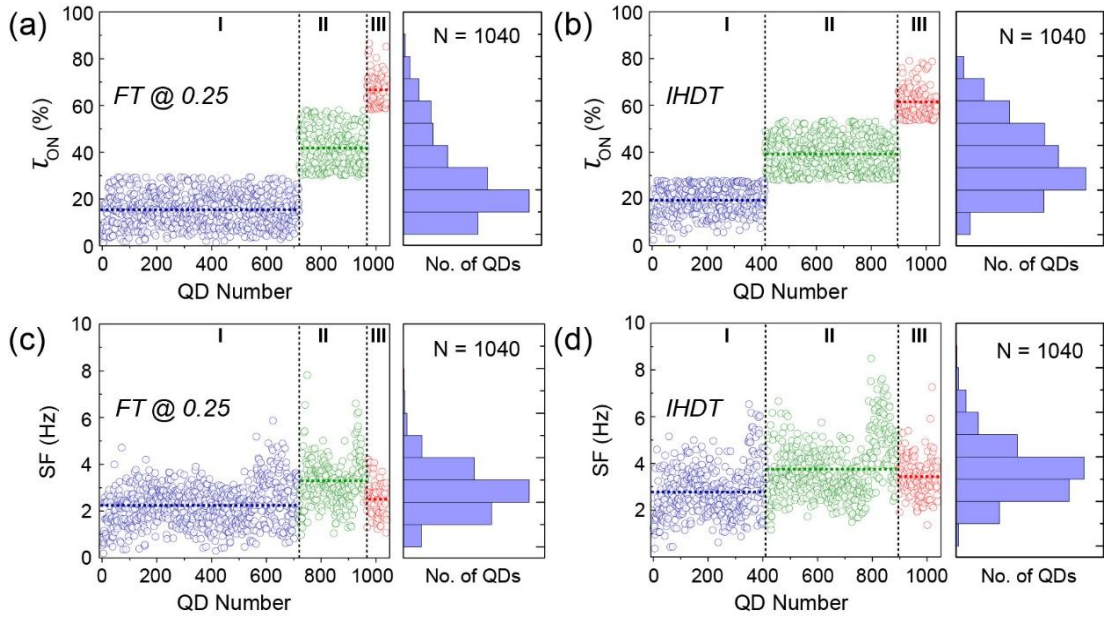

**Figure S6.** Category-wise distributions of $\tau_{ON}$ (%) from 1040 $Mn^{+2}$ doped ZnCdS single emitters using fixed threshold (*FT*) (top left, (a)) and *IHDT* (top right, (b)) blinking analysis technique with their sub-ensemble mean values. Class-wise distribution of switching frequency (*SF*, in Hz) for the same emitters from *FT* (bottom left, (c)) and *IHDT* (bottom right, (d)) method. *Blue*, *green* and *red* color code stands for corresponding sub-ensembles *"Mostly Off"*, *"Intermediate"* and *"Mostly On"* respectively.

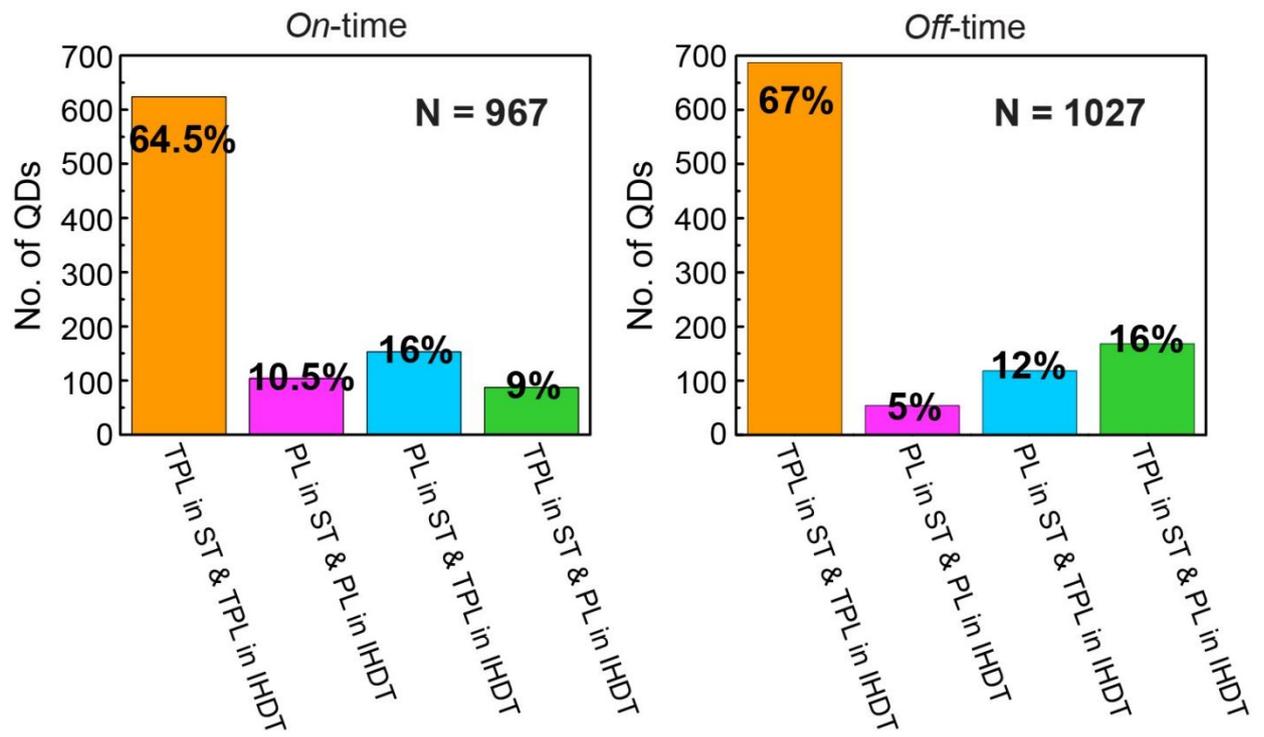

**Figure S7.** Shows the percentages of emitters which changes the characteristic nature (power law (*PL*)/ truncated power law (*TPL*)) of their *On-/Off*-time distribution ($P(t_{On/Off})$) with changing blinking analysis method from single fixed threshold (*FT*) to *IHDT*.

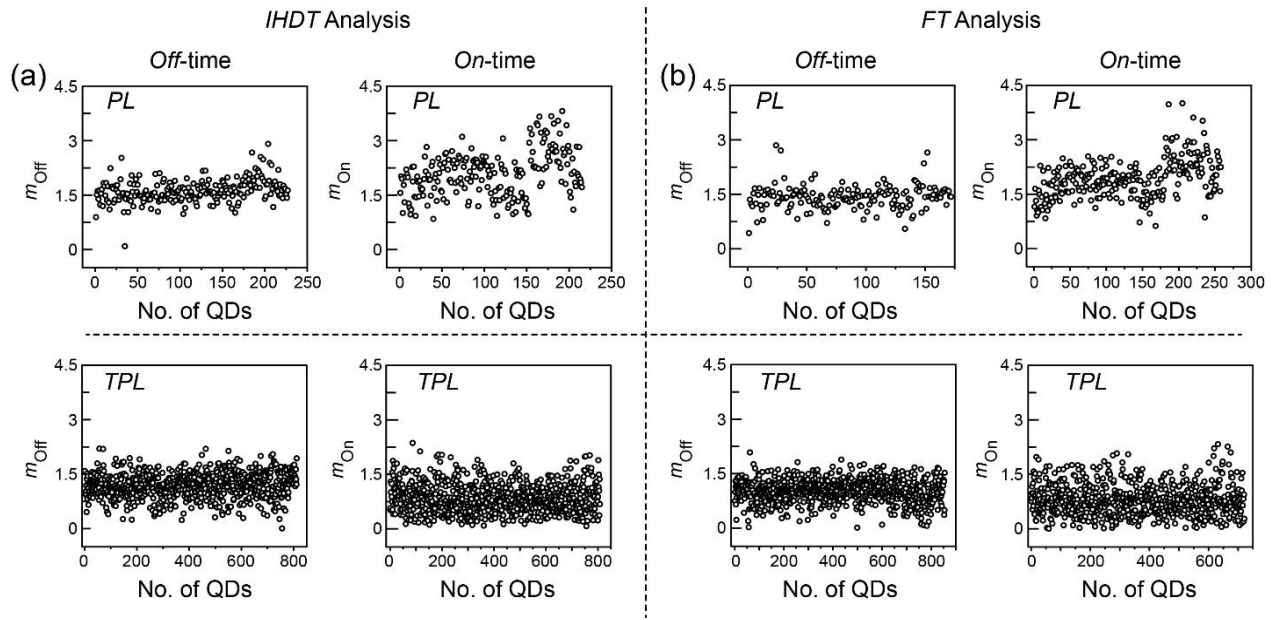

**Figure S8.** Distribution of power law (*PL*) (upper panel) and truncated power law (*TPL*) (lower panel) exponent values for $P(t_{On/Off})$ in *IHDT* blinking analysis technique (a). (b) Distributions of the same have been represented in (b), using *FT* analysis @ 0.25.

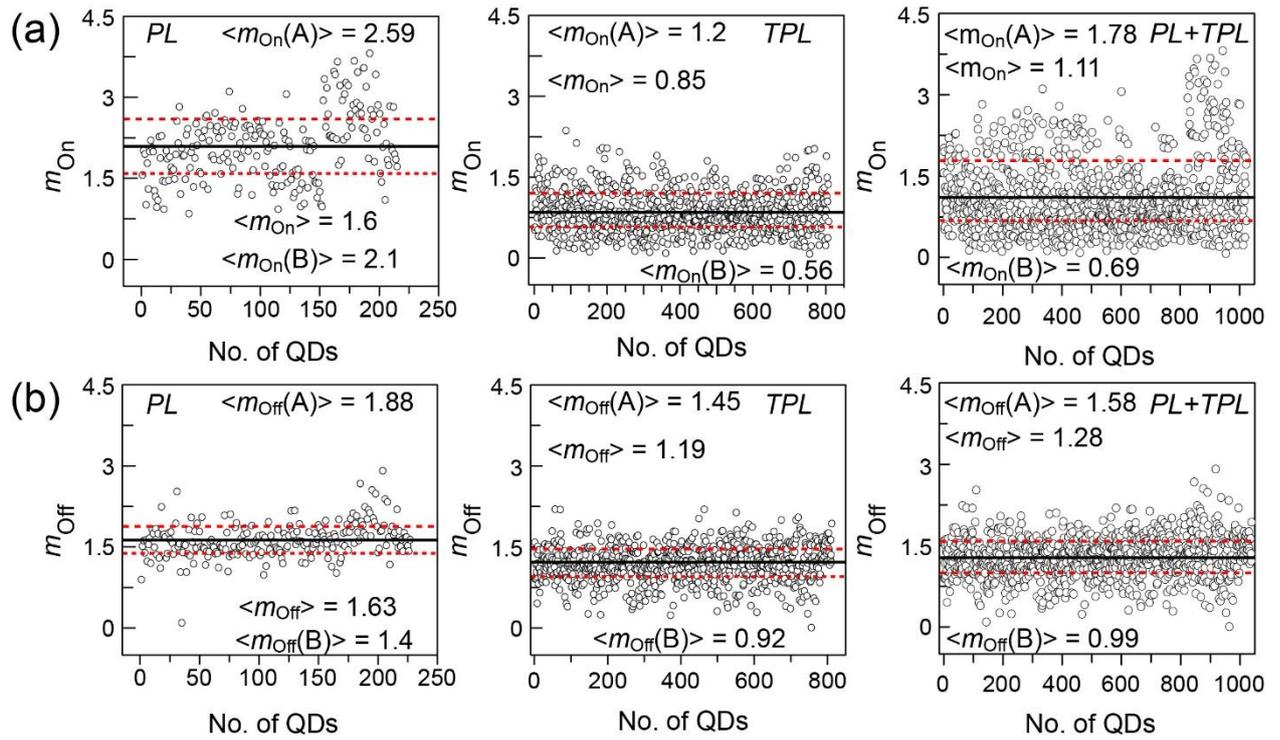

**Figure S9.** Comparison of the average of the exponent values, lying *above* the ensemble mean ($\langle m_{On/Off}(A)\rangle$, red dotted line) and mean of the exponent value *below* to the ensemble mean ($\langle m_{On/Off}(B)\rangle$, red dotted line) with the ensemble average exponent values ($\langle m_{On/Off}(PL/TPL)\rangle$, black solid line). Situations considering the *On*- events are shown in the *upper* panel (a) while the same for *Off*- events are depicted in the *lower* panel (b).

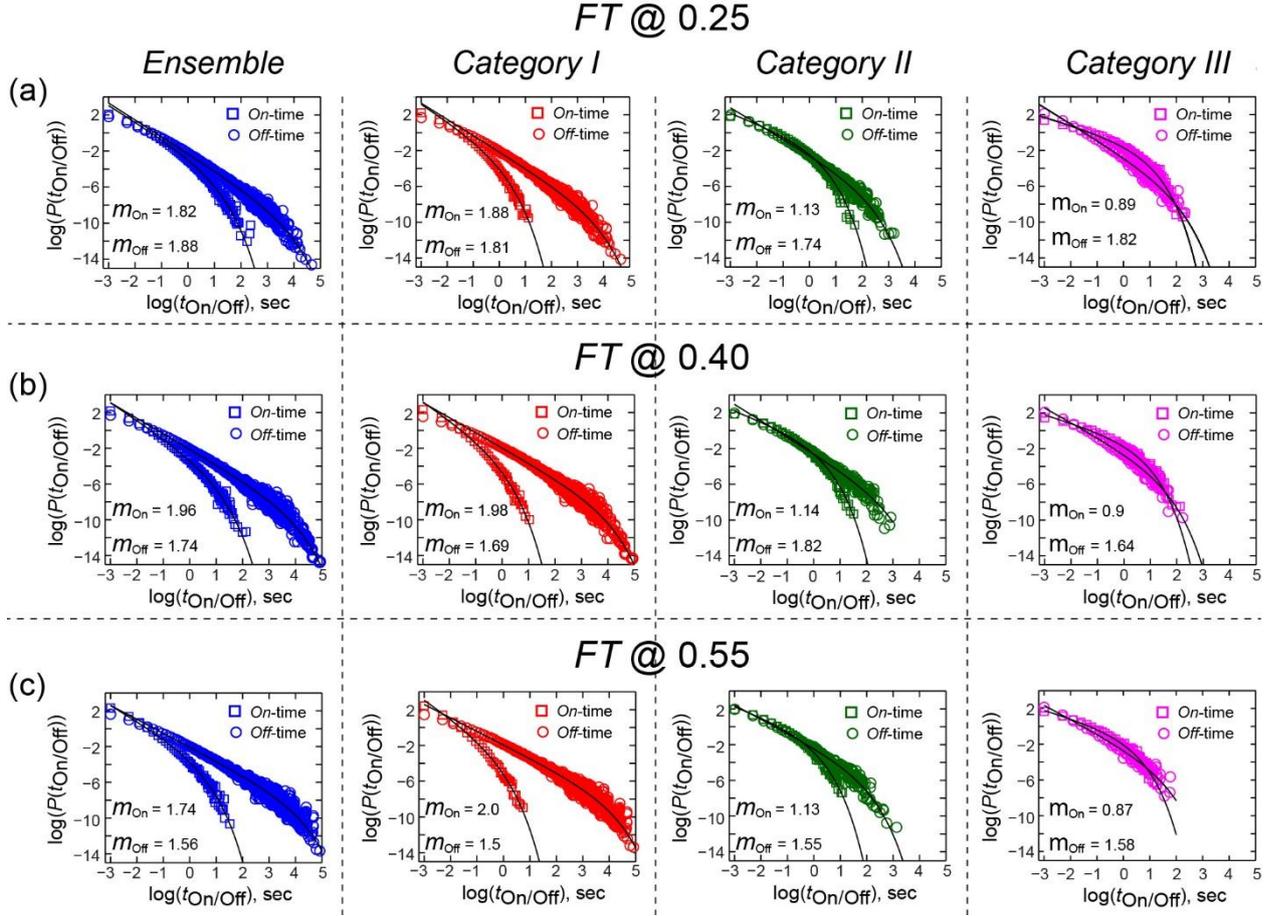

**Figure S10.** Time-ensemble averaged $\langle\overline{P(t_{On/Off})}\rangle$ from the ensemble and sub-categories of the QDs, using *FT* method @ 0.25 (a), 0.4 (b) & 0.55 (c). It is noted that the exponent ($\langle\overline{m_{On/Off}}\rangle$) values from time-ensemble averaged $\langle\overline{P(t_{On/Off})}\rangle$ are not comparable within the (sub)ensembles in each of the cases.

# Supplementary Data (Tables)

**Table SI**. The number (or %) of QDs which changes (or does not change) the characteristic of $P(t_{On/Off})$ nature (power law to truncated power law or *vice versa*) with the method of analysis

| $P(t_{On/Off})$ | Nature of $P(t_{On/Off})$ for FT@0.25 | Nature of $P(t_{On/Off})$ for IHDT | Number of QDs (% of ensemble) |
|---|---|---|---|
| $P(t_{On})$ | PL | PL | **103 (10.5)** |
|  | PL | TPL | **153 (16)** |
|  | TPL | PL | **87 (9)** |
|  | TPL | TPL | **624 (64.5)** |
| $P(t_{Off})$ | PL | PL | **54 (5)** |
|  | PL | TPL | **118 (12)** |
|  | TPL | PL | **168 (16)** |
|  | TPL | TPL | **687 (67)** |

**Table SII.** (Sub)ensemble mean values and standard deviations of different blinking parameters for an ensemble (1040) of $Mn^{+2}$ doped ZnCdS QDs, using *IHDT* and *FT* (@ 0.25):

| Blinking Parameters | | Category I + II + III | | Category I | | Category II | | Category III | |
|---|---|---|---|---|---|---|---|---|---|
| | | IHDT | FT @0.25 | IHDT | FT @0.25 | IHDT | FT @0.25 | IHDT | FT @0.25 |
| $\langle \tau_{ON}(\%) \rangle$ | | 34.40 (± 15.64) | 25.42 (± 17.58) | 19.41 (± 6.00) | 15.5 (± 6.98) | 39.15 (± 7.23) | 42 (± 8.38) | 61.34 (± 6.42) | 66.6 (± 7.0) |
| $\langle SF \rangle$ (In Hz) | | 3.33 (± 1.21) | 2.50 (± 1.02) | 2.79 (± 1.08) | 2.24 (± 0.89) | 3.76 (± 1.22) | 3.28 (± 1.03) | 3.44 (± 0.85) | 2.52 (± 0.76) |
| (Sub) Ensemble average $P(t_{On})$ parameters | $\langle m_{On} \rangle$ (PL+TPL) | 1.11 (± 0.68) | 1.09 (± 0.67) | 1.40 (± 0.74) | 1.27 (± 0.71) | 0.98 (± 0.60) | 0.75 (± 0.36) | 0.72 (± 0.28) | 0.66 (± 0.33) |
| | $\langle m_{On} \rangle$ (PL) | 2.09 (± 0.62) | 1.93 (± 0.53) | 2.11 (± 0.56) | 1.98 (± 0.5) | 2.14 (± 0.70) | 1.42 (± 0.43) | 1.15 (± 0.23) | 1.10 (± 0.4) |
| | % PL | ~21 | ~26 | ~37 | ~36 | ~12 | ~6 | ~5 | ~8 |
| | $\langle m_{On} \rangle$ (TPL) | 0.85 (± 0.40) | 0.8 (± 0.41) | 1.0 (± 0.47) | 0.87 (± 0.46) | 0.81 (± 0.36) | 0.71 (± 0.31) | 0.7 (± 0.27) | 0.62 (± 0.29) |
| | $\langle \tau_{c(On)} \rangle$ | 0.61 (± 0.68) | 0.58 (± 0.79) | 0.56 (± 0.76) | 0.45 (± 0.6) | 0.6 (± 0.65) | 0.61 (± 0.69) | 0.75 (± 0.60) | 1.3 (± 1.49) |
| | % TPL | ~79 | ~74 | ~63 | ~64 | ~88 | ~94 | ~95 | ~92 |
| (Sub) Ensemble average $P(t_{Off})$ parameters | $\langle m_{Off} \rangle$ (PL + TPL) | 1.28 (± 0.38) | 1.08 (± 0.35) | 1.14 (± 0.32) | 1.02 (± 0.33) | 1.31 (± 0.36) | 1.19 (± 0.31) | 1.61 (± 0.37) | 1.36 (± 0.45) |
| | $\langle m_{Off} \rangle$ (PL) | 1.63 (± 0.32) | 1.42 (± 0.33) | 1.4 (± 0.27) | 1.34 (± 0.29) | 1.66 (± 0.29) | 1.53 (± 0.30) | 1.86 (± 0.25) | 1.67 (± 0.4) |
| | %PL | ~22 | ~17 | ~17 | ~16 | ~22 | ~13 | ~36 | ~31 |
| | $\langle m_{Off} \rangle$ (TPL) | 1.19 (± 0.33) | 1.02 (± 0.31) | 1.08 (± 0.30) | 0.96 (± 0.30) | 1.21 (± 0.31) | 1.14 (± 0.27) | 1.47 (± 0.36) | 1.22 (± 0.41) |
| | $\langle \tau_{c(Off)} \rangle$ | 1.32 (± 1.38) | 1.38 (± 1.50) | 1.39 (± 1.34) | 1.46 (± 0.51) | 1.2 (± 1.28) | 1.21 (± 1.53) | 1.58 (± 1.78) | 1.11 (± 1.22) |
| | % TPL | ~78 | ~83 | ~83 | ~84 | ~78 | ~87 | ~64 | ~69 |

# Supplementary References

# Supplementary Movie

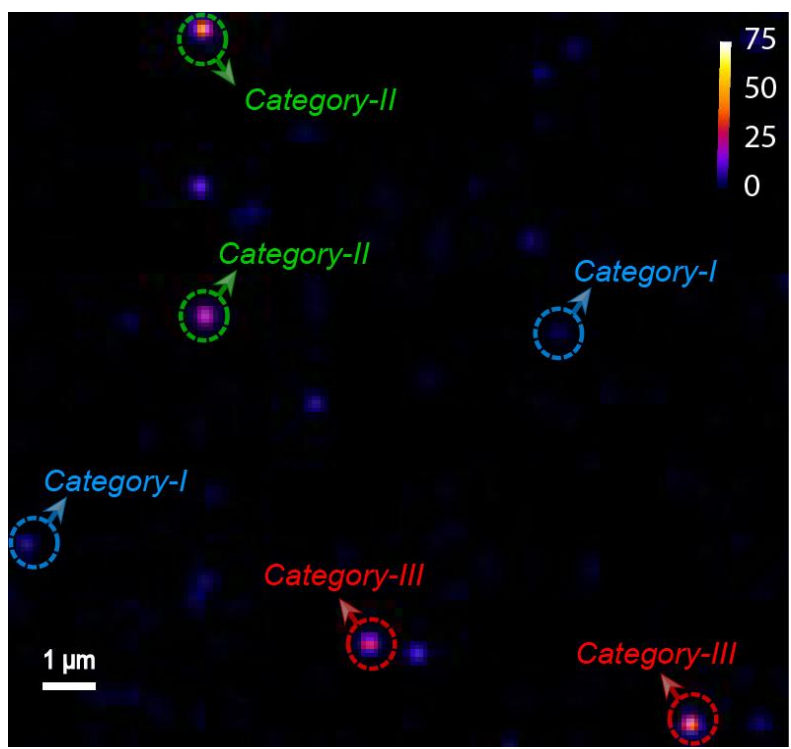

**Movie M1.** A photoluminescence intensity movie depicting diverse blinking characteristics of several individual $Mn^{+2}$ doped ZnCdS NCs, immobilized in PMMA on a glass substrate. The samples were excited at 457 nm (500 $Wcm^{-2}$) and the movie was acquired at 20 Hz, through a 545-635 nm band-pass filter and imaged using a CCD camera.